\documentclass[lettersize,journal]{IEEEtran}
\usepackage{amsmath,amsfonts}
\usepackage{algorithmic}
\usepackage{algorithm}
\usepackage{array}
\usepackage[caption=false,font=normalsize,labelfont=sf,textfont=sf]{subfig}
\usepackage{textcomp}
\usepackage{stfloats}
\usepackage{url}
\usepackage{verbatim}
\usepackage{graphicx}
\usepackage{cite}
\newtheorem{definition}{Definition}[section]
\usepackage{multirow}
\usepackage{orcidlink}
\usepackage{booktabs}
\newcommand{\jy}[1]{\textcolor{black}{#1}}
\usepackage{hyperref}
\usepackage{breakurl}
\begin{document}

\title{Computational Performance Bounds Prediction in Quantum Computing with Unstable Noise}

\author{Jinyang Li, Samudra Dasgupta, Yuhong Song, Lei Yang, Travis Humble, Weiwen Jiang

\thanks{This work was supported in part by the NSF 2311949, 2320957, 2440637, 2513431. This work was also supported by the U.S. Department of Energy, Office of Science, Office of Advanced Scientific Computing Research through the Accelerated Research in Quantum Computing Program MACH-Q Project. The research used IBM Quantum resources via the Oak Ridge Leadership Computing Facility at the Oak Ridge National Laboratory, which is supported by the Office of Science of the U.S. Department of Energy under Contract No. DE-AC05-00OR22725.}
\thanks{Jinyang Li, Yuhong Song, and Weiwen Jiang are with the Department of Electrical and Computer Engineering
at George Mason University and Quantum Science \& Engineering Center, Fairfax, VA 22030 USA (e-mail: jli56@gmu.edu, {ysong34@gmu.edu}, wjiang8@gmu.edu). Lei Yang is with the Department of Information Sciences and Technology at George Mason University, Fairfax, VA 22030 USA (e-mail: {lyang29@gmu.edu}).}
\thanks{Samudra Dasgupta and Travis Humble are with the Quantum Science Center, Oak Ridge National Laboratory, Oak Ridge, TN 37831 USA (e-mail: samudra.dasgupta@gmail.com, humblets@ornl.gov).}
\thanks{© 2025 IEEE.  Personal use of this material is permitted.  Permission from IEEE must be obtained for all other uses, in any current or future media, including reprinting/republishing this material for advertising or promotional purposes, creating new collective works, for resale or redistribution to servers or lists, or reuse of any copyrighted component of this work in other works.}
}




\maketitle

\begin{abstract}
Quantum computing has significantly advanced in recent years, boasting devices with hundreds of quantum bits (qubits), hinting at its potential quantum advantage over classical computing. 
Yet, noise in quantum devices poses significant barriers to realizing this supremacy. Understanding noise's impact is crucial for reproducibility and application reuse; moreover, the next-generation quantum-centric supercomputing essentially requires efficient and accurate noise characterization to support system management (e.g., job scheduling), where ensuring correct functional performance (i.e., fidelity) of jobs on available quantum devices can even be higher-priority than traditional objectives.
However, noise fluctuates over time, even on the same quantum device, which makes predicting the computational bounds for on-the-fly noise is vital.
Noisy quantum simulation can offer insights but faces efficiency and scalability issues. 
In this work, we propose a data-driven workflow, namely \textit{QuBound}, to predict computational performance bounds. It decomposes historical performance traces to isolate noise sources and devises a novel encoder to embed circuit and noise information processed by a Long Short-Term Memory (LSTM) network. 
For evaluation, we compare \textit{QuBound} with a state-of-the-art learning-based predictor, which only generates a single performance value instead of a bound.
Experimental results show that the result of the existing approach falls outside of performance bounds, while all predictions from our \textit{QuBound} with the assistance of performance decomposition better fit the bounds.
Moreover, \textit{QuBound} can efficiently produce practical bounds for various circuits with over $10^6$ speedup over simulation; in addition, the range from \textit{QuBound} is over $10\times$ narrower than the state-of-the-art analytical approach.
\end{abstract}


\section{Introduction}
\IEEEPARstart{W}{e} are now witnessing the rapid growth of quantum hardware (e.g., IBM quantum scaled from 5 qubits in 2016 to 1,121 qubits in 2023), which raises the great potential for next-generation quantum-centric supercomputing \cite{gambetta2022quantum, britt2017high, alexeev2023quantum, george2023quantum, bravyi2022future}, where multiple quantum processors will interplay with high-performance computing (HPC) systems, aiming to significantly boost computational capacity for real-world applications, such as chemistry and finance \cite{sajjan2022quantum, lee2023evaluating, singh2023benchmarking, herman2023quantum, leclerc2023financial, jiang2021co, li2023novel,li2024quapprox}.
However, the practical use of these quantum processors faces notable challenges, especially the inherent noise in quantum devices, which are prone to high errors.
Due to multiple noise sources exhibiting randomness, such as thermal noise related to the ambient computing environment, noise on quantum devices presents instability and unpredictability.
As a result, the performance of an application will fluctuate over time \cite{dasgupta2020characterizing}, \jy{and several works have explored methods to handle noise in order to improve real-world performance \cite{hu2023battle,he2024distributionally}.}

Unlike classical computing\cite{yang2025novel,yang2023device}, the system management (e.g., job scheduling) for quantum-centric computing\cite{li2025qusplit} needs to consider not only the traditional objectives (e.g., resource utilization and job wait time) but also the functional correctness (i.e., fidelity) of an application on noisy quantum processors.
Therefore, it calls for fidelity-aware system management.
In this context, the accurate and efficient performance prediction of quantum applications becomes essential, which provides the metric for management optimizations.
This work aims to develop an efficient workflow for performance-bound prediction of quantum circuits at runtime.
\jy{Such a capability is particularly valuable for system schedulers selecting execution backends and for compiler developers evaluating transpilation or circuit layout strategies under time-varying noise conditions.}

One straightforward solution is to run a noisy simulation to capture the performance bounds \cite{ravi2021adaptive}; however,
it lacks efficiency in two aspects.
First, noisy simulations are time-consuming, but noise changes frequently; thus, the obtained bounds might already be outdated after the simulation.
Second, the simulation faces scalability issues; the memory requirements of a full-state simulation grow exponentially along with the number of qubits, easily exceeding the memory capacity on classical computers.
Another solution is to carry out statistic analysis; a recent research work \cite{dasgupta2023reliability} proposes to predict the theoretical performance bounds by calculating the distance of the distributions under different noise. It can achieve significant speedup at runtime; however, the theoretic analysis is commonly coupled with worst-case analysis, which leads to a too-loose predicted bound.
To make the performance prediction practical, we aim to minimize the prediction time and bound range, while maximizing the ratio of successfully predicted run-time performance between the bounds.






In this paper, we rethink the solution from a data-driven perspective.
As cloud access to quantum computers became available in 2016, and the quantum providers have already collected and released enough quantum noise data, this provides the potential to use data-driven approaches for performance predictions.
While it's motivating, challenges remain. 
First, a mismatch exists between performance and noise sources.
Specifically, all quantum noise sources contribute to the performance, but we can only get the performance as a whole.
Moreover, the off-the-shelf noise model only covers part of noise sources, saying it provides device-related noise information but lacks sampling noise information, which comes from the probabilistic nature of quantum mechanics.
As such, without decomposing the performance and matching it to the related noise sources, it's hard to accurately predict the performance.
Second, even if we can decompose and match the performance with noise sources, 
the appropriate machine learning (ML) architecture for performance prediction is unclear. More importantly, it's unknown how to encode noise with a quantum circuit and process it by the ML algorithm.








This paper proposes a data-driven performance-bound prediction framework, namely \textit{QuBound}.
It is a dual-component workflow that combines performance decomposition with an ML-based performance predictor. The first component, \textit{QuDECOM}, focuses on decomposing the performance of quantum circuits into trend and residual parts, isolating device noise from sampling noise. This decomposition not only enhances the accuracy of our predictor by using the pair of device noise and trend performance, but also enables us to predict upper and lower bounds by using the residual performance. 
The second component, \textit{QuPRED}, contains a novel encoding strategy that integrates detailed quantum circuit information with noise characteristics, capturing the interplay between quantum operations and noise effects. 
Then, \textit{QuBound} applies the Long Short-Term Memory (LSTM) model, renowned for its ability to handle sequential data, corresponding to the time-series historic noise and performance traces in our work.
The model then processes the encoded information to predict the performance of a quantum circuit under specific noise.


The main contributions of this paper are as follows:
\begin{itemize}
    \item We formally define the quantum bound prediction problem, denoted ${QuBound_p}$.
    \item \textit{QuBound} innovatively proposes a performance decomposition approach \textit{QuDECOM} to isolate the effects of different types of noise on performance. 
    \item To the best of our knowledge, the solver \textit{QuBound} is the very first data-driven workflow to combine ML and performance decomposition to predict the performance bounds on quantum computers with unstable noise.
    \item A dataset is built with pairs of noise sources and decomposed performance, on top of which, a novel data encoder can create embeddings of noise data and quantum circuits, enabling an LSTM-based performance prediction.
\end{itemize}

We evaluate \textit{QuBound} on various circuit types. We first compare it with learning-based methods that predict only a single value. Results show that \textit{QuBound} more accurately estimates performance within the ideal bound. We further compare it with two bound-generating methods: (1) repeated noisy simulations and (2) theoretical analysis. \textit{QuBound} is evaluated in terms of bound range, runtime, and prediction accuracy under different noise conditions. It generates tight, practical bounds with low latency on both small (e.g., GHZ-3) and large circuits (e.g., GHZ-15).
We also demonstrate the effectiveness of \textit{QuBound} across multiple quantum backends.

The paper is organized as follows. Section \ref{section2} discusses the background and related works. Section \ref{section3} defines the problem. Section \ref{section4} reveals the motivation and challenges. Section \ref{section5} details the proposed workflow \textit{QuBound}. Experimental results and discussions are given in Section \ref{section6} and Section \ref{section7}. Last, conclusion remarks are in Section \ref{section8}.



\section{Background and Related Work}\label{relatedwork}
\label{section2}
\begin{figure}[t]
    \centering
    \includegraphics[width=0.48\textwidth]{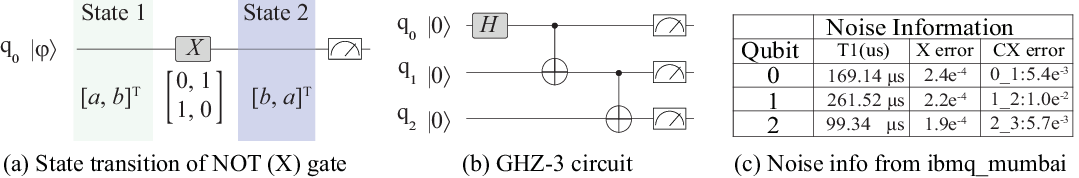}
    \caption{Example of quantum computing, quantum circuit, and quantum noise.}
    \label{fig:circuitexample}
    \vspace{-10pt}
\end{figure}


Quantum computing uses quantum bits (qubits) to express information. The state of a qubit $q_0$ can be represented as \(|q_0\rangle = a|0\rangle + b|1\rangle\) or a state vector $[a,b]^{T}$, as shown in Figure \ref{fig:circuitexample}(a), where \(a\) and \(b\) are complex numbers, which respectively represent the amplitudes of the qubit $q_0$ being in the \(|0\rangle\) or \(|1\rangle\) state.
Correspondingly, $|a|^2$ and $|b|^2$ are the probabilities of obtaining these states if a readout measurement is performed.
The state probabilities must sum to $1$ (i.e., \(|a|^2 + |b|^2 = 1\)).



Noise on quantum devices is unstable; that is, noise parameters (e.g., T1/T2 time and gate error) are dynamically changed over time.
This can be caused by various reasons, such as the uncertain environment, the operational status of quantum devices, and the interaction between qubits and their surroundings. 
The impact of unstable noise on quantum application performance is significant \cite{dasgupta2022characterizing, hu2023battle, dasgupta2022assessing, dasgupta2021stability, zhang2023disq, wang2023dgr}.
For example, work \cite{hu2023battle} demonstrates that the circuit accuracy for the earthquake detection task can vary from 50\% to over 80\% under different noise conditions.

Due to the presence of high-level and unstable noise in quantum devices, fidelity is one of the most important metrics for quantum performance.
Given the probability of obtaining the $i^{th}$ outcome of a quantum state in an ideal environment as \( p_{\text{ideal}, i} \), and that in a noisy environment as \( p_{\text{observed}, i} \), the fidelity can be quantified by the equation:
\begin{equation}
    F = \left( \sum_i \sqrt{p_{\text{ideal}, i} \cdot p_{\text{observed}, i}} \right)^2.
\end{equation}



Fidelity prediction is the key to system optimization, such as circuit compilation and resource management in quantum-centric high-performance computing systems. 
Three main approaches exist for predicting the performance of quantum circuits.
First, the simulation-based prediction method uses noise models to simulate quantum circuits multiple times to get the estimated results. In quantum resource management works \cite{ravi2021adaptive,ohkura2022simultaneous,niu2022parallel}, work \cite{ravi2021adaptive} applies the noisy simulator to get the fidelity, which is
time-consuming and not scalable.

Second, there exist statistical methods to calculate the fidelity \cite{yu2022statistical, liu2020reliability}. However, these 
pioneer works do not consider run-time noise. In compilation works, noise-aware fidelity is calculated by a simple equation as a metric to optimize the compilation
\cite{nishio2020extracting}; however, this is inaccurate since the quantum circuit structure is not considered. Recently, work \cite{dasgupta2023reliability} uses the analysis of noise distribution to calculate the fidelity.

Last, another research direction is to apply machine learning (ML)-based methods for fidelity prediction \cite{zhang2021direct,vadali2024quantum,wang2022quest, saravanan2022data}.
Work \cite{wang2022quest} predicts fidelity using the graph transformer to encode quantum circuits and noise. 
A more detailed discussion on the problem of existing work will be provided in section \ref{section4}.

\section{Problem Formulation}
\label{section3}

\begin{figure}[t]
	\centering
	\includegraphics[width=0.38\textwidth]{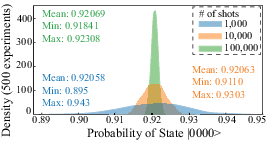}
	\caption{Performance impacted by incomplete measurements with noise $N$.}
	\vspace{-8pt}
	\label{fig:incomplete_meas}
\end{figure}
\subsection{Quantum Circuit}
A quantum circuit is a sequence of quantum gates (operations) that manipulate qubits to perform computation. 
In this paper, we represent a quantum circuit \( C = \langle Q, G \rangle \), where \( Q = \langle q_0, q_1, \ldots, q_m \rangle \) is a set of qubits and \( m \) is the index of the qubit from top to bottom in the circuit. \( G = \langle g_0, g_1, \ldots, g_n \rangle \) is a set of quantum gates. For each quantum gate $g$  here, it should be in two forms, either $X_{a,s}$ or $X_{b,c, s}$ where \( X \) represents the name of the quantum gate. The subscript \( a \) denotes the index of the qubit $q_a\in Q$ that the single-qubit gate acts on, while $X_{b,c}$ represents a two-qubit gate, and the subscripts \( b, c \) denote the indices of the qubits that the two-qubit gate acts on. Then, the subscript \( s \) denotes the step/index of a quantum gate on one qubit.
For a 3-qubit Greenberger-Horne-Zeilinger (GHZ) circuit, as shown in Figure \ref{fig:circuitexample}(b),
it can be represented as \( C_{GHZ3} = \langle Q, G \rangle \), where \( Q = \langle q_0, q_1, q_2 \rangle \) and \( G = \langle H_{0,0}, CNOT_{0,1,0}, CNOT_{1,2,1} \rangle \).

\subsection{Quantum Noise}\label{sec:qnoise}

\textbf{Noise Source.}
Quantum noise is inherent in quantum computers and is typically characterized by several key parameters:
\textit{1) T1 relaxation time}, which is the \jy{time taken for a qubit} to relax from the excited state \(|1\rangle\) to the ground state \(|0\rangle\). 
\textit{2) T2 dephasing time}, which quantifies how quickly superposition states, such as \(|+\rangle\) or \(|-\rangle\), lose their phase information, leading to a decay in qubit coherence. 
\textit{3) Gate errors}, which are associated with quantum gates, occur when the quantum gates fail to perform their intended operations.
Figure \ref{fig:circuitexample}(c) shows an example of historical noise information from a real quantum device.
Given a quantum circuit \( C = \langle Q, G \rangle \) for qubit $q_i\in Q$ and gate $g_j\in G$, the noise of executing gate $g_j$ at step $s$ on $q_i$ can be represented as $N(q_i,g_j,s)$ which can be computed as follows.
The noise for a qubit-gate pair is defined as:
\begin{equation}
N(q_i, g_j) = \begin{cases} 
n_{\text{gate}}(q_i, g_j)  & \text{if } g_j \text{ is a 1-qubit gate} \\
n_{\text{gate}}(q_i, q_k, g_j) & \text{if } g_j \text{ is a 2-qubit gate} 
\end{cases}
\end{equation}
where
\( n_{\text{gate}}(q_i, g_j) \) is the gate error for a single-qubit gate \( g_j \) acting on qubit \( q_i \),  and \( n_{\text{gate}}(q_i, q_k, g_j) \) is the gate error  for a two-qubit gate \( g_j \) acting between qubits \( q_i \) and \( q_k \).

Thus, the overall noise model for the quantum circuit is represented by the set:
\begin{equation}
N = \left\{ N(q_i, g_j) \cup N_{\text{T1}}(q_i) \cup N_{\text{T2}}(q_i) \mid q_i \in Q, g_j \in G \right\}
\end{equation}
where
\( N_{\text{T1}}(q_i) \) is the relaxation time (T1) for qubit \( q_i \),  and \( N_{\text{T2}}(q_i) \) is the dephasing time (T2) for qubit \( q_i \).


\textbf{Unstable Quantum Noise.}
To formulate the unstable noise, we define the ``noise trace", denoted as $trN$, as the collection of quantum noise parameters from a quantum device over a specified time period.
The noise trace $trN$ is represented as:
\begin{equation}
trN = \left\{ N_t \mid t \in [t_0, t_f] \right\}
\end{equation}
where \( N_t \) is the noise model at time \( t \), and \( [t_0, t_f] \) represents the time interval over which the noise data is collected. Each \( N_t \) can be defined as:
\begin{equation}
N_t = \left\{ N(q_i, g_j, t) \mid q_i \in Q, g_j \in G \right\}
\end{equation}
Here, \( N(q_i, g_j, t) \) denotes the noise \( N(q_i, g_j) \) at time \( t \).


\subsection{Quantum Performance}
\label{sec:quantumperformance}
\textbf{Performance Trace.}
The fidelity defined in Section \ref{relatedwork} is one kind of performance; however, it is quite expensive to obtain it on a measurement-based quantum computer, as it requires the amplitude of each quantum state in a perfect environment. 
Instead, we use these two measurement-friendly metrics.
(1) \textit{Probability-based}, which evaluates the probability of obtaining specific output states (e.g., the probability of \(|00\rangle\) in a 2-qubit system is \(|a|^2\)); and 
(2) \textit{Observable-based}, which computes the expectation value of a given observable, such as Pauli operators. For example, measuring \(Z \otimes Z\) yields \(P = \langle \psi | ZZ | \psi \rangle = |a|^2 + |d|^2 - |b|^2 - |c|^2\), often used in quantum chemistry for energy estimation.


Given a quantum circuit $C$, and the noise of a quantum system at time $t$ is $N_t$, the performance $P_t$ is defined as below:
\begin{equation}
P_t = f(C, N_t)
\end{equation}
where $f$ is a selection function representing the probability- or observable-based measurements.


We define the ``performance trace", denoted as $trP$, as the collection of performance measurements with a specific quantum circuit when subjected to the noise trace $trN$.
Formally, the performance trace $trP$ can be defined as a time series:
\begin{equation}
trP = \left\{ P_t \mid t \in [t_0, t_f] \right\}
\end{equation}
where \( P_t \) denotes the quantum circuit performance at time \( t \).


\textbf{Performance Bound.}
The quantum performance is mainly affected by two sources: \textit{i)} noise $N$ by hardware defects and environmental interactions, which has been discussed in Section \ref{sec:qnoise}; and \textit{ii)} incomplete measurements (i.e., finite number of shots).
In practice, the quantum processors have limited shots per task; for example, IonQ's Aria, QuEra, IonQ's Harmony, and IBM are limited by 5,000, 1,000, 10,000, and 100,000 shots, respectively \cite{Amazon_Quotas,IBM_Quantum_Shots}; moreover, money is charged per shot (e.g., \$0.01 per shot by AWS Braket). 
We use an experiment to test the effects of incomplete measurements on a 4-qubit variational quantum eigensolver (VQE) circuit, as shown in Figure \ref{fig:incomplete_meas}.
The results draw us a crucial insight: Even with the same quantum noise, the incomplete measurement will lead to different performance. This indicates that \textit{the performance predictor that only takes the device-related noise model as inputs should \underline{estimate a practical pair of bounds} instead of a single fidelity value}.

In this context, given a quantum circuit $C$, a noise at time $t$ as $N_t$, a shot number $s$, and a confidence level $CL$, the pair of performance upper bound and lower bound is defined as $\langle P_{low,s,CL}(C,N_t),P_{up,s,CL}(C,N_t)\rangle$, such that $prob\{P_{low,s,CL}\le P_t\le P_{up,s,CL}(C,N_t)\}\ge CL$ and $prob\{P_t\le P_{low,s,CL}(C,N_t)\}=prob\{P_t\ge P_{upper,s,CL}(C,N_t)\}$.
Then, we define the Bound Range, $Range_{s,CL}$ as below:
\begin{equation}
    Range_{s,CL} = P_{up,s,CL}(C,N_t)-P_{low,s,CL}(C,N_t)
\end{equation}



\textbf{Bound-Compliance Rate.}
To evaluate the accuracy and effectiveness of the performance bounds, we define an additional metric called the `Bound-Compliance Rate ($BCR$)'. This metric represents the proportion of runtime performance measurements that successfully fall within the predicted computational performance bounds. 
The $BCR$ is defined as:
\begin{equation}
\text{$BCR$} = \frac{\text{Number of measurements within bounds}}{\text{Total number of measurements}} 
\end{equation}
A higher $BCR$ \jy{indicates more reliable prediction bounds} of the circuit's performance.

\subsection{Problem Definition}


With the above definition, we formally define the problem, namely $QuBound_p$, as below.

\begin{definition}\label{def:prob}
Given a quantum circuit $C$, shot number $s$, the current noise $N_t$ and the required confidence level $CL$, $QuBound_p$ is to determine:
\begin{enumerate}
    \item The upper performance bound $P_{up,s,CL}(C,N_t)$,
    \item The lower performance bound $P_{low,s,CL}(C,N_t)$,
\end{enumerate}
such that the $Range_{s,CL}$ is minimized while the $BCR$ is maximized. 
The historical performance trace $trP$ and the noise trace $trN$ are available for use to expedite the prediction.
\end{definition}

\section{Motivation and Challenges}
\label{section4}
\begin{figure*}[t]
    \centering
    \includegraphics[width=1\textwidth]{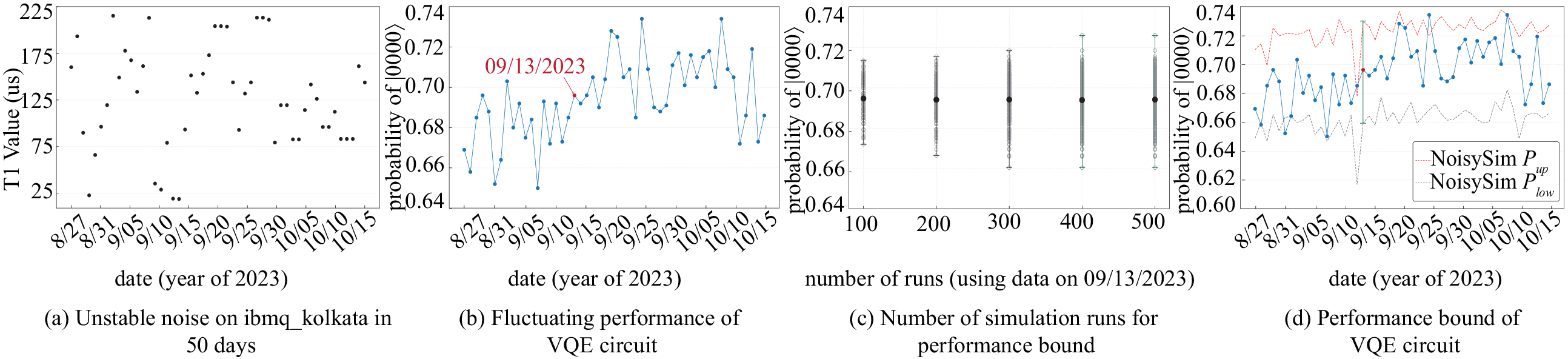}
    \caption{Illustration of the unstable noise, the unstable performance of the circuit, and the limitation of using noisy simulations to obtain predictions.}
    \label{fig:mot1}
    \vspace{-10pt}
\end{figure*}


\subsection{Observation and Motivation}

\textbf{Observation: The presence of unstable noise leads to the unpredictable performance of quantum circuits.}

Figure \ref{fig:mot1}(a) shows the daily change of $T1$ (i.e., thermal relaxation time), on IBM \verb|ibmq_kolkata| quantum computers in the past two months.
$T1$ is one of the contributing factors to quantum system decoherence, which introduces noise-like effects in the system. 
We can see from Figure \ref{fig:mot1}(a) that $T1$ changes over time which can be up to 200 $\mu s$ and lower under 25 $\mu s$.
As pointed out by recent studies \cite{dasgupta2020characterizing,hu2023battle}, other noise-related metrics, such as $T2$, gate error, readout error, etc., have similar behavior.
Moreover, the probabilistic nature of quantum mechanics, along with being inherently sensitive to the running environment, leads to the unpredictability and apparent randomness associated with quantum noise.

Such a noisy, unstable, and unpredictable running environment will lead to fluctuating performance.
We examine this by repeatedly executing one VQE circuit --- the quantum circuit widely used in quantum chemistry and materials science --- on backend \verb|ibmq_kolkata|.
Results in 
Figure \ref{fig:mot1}(b) show that the 
circuit outcome fluctuates heavily.
Specifically, the 
first quantum state $|0000\rangle$ ranges from 0.66 to nearly 0.74. 
This observation shows the unpredictability of the quantum circuit simulation result under different noise parameters.

\textbf{Motivation: A need to assess and characterize result variability under unpredictable quantum noise.}

The unstable performance of actual quantum computers raises serious issues in resource management in the next-generation quantum-centric supercomputing, where the incoming jobs need to be allocated to a pool of quantum processors for execution.
Specifically, jobs' fluctuating performance will change the optimal resource allocation strategies. 
The need for fidelity-aware optimization motivates this work to predict the bound of performance for a given circuit using the noise information obtained at run-time.
A good predicted bound with a practical range can provide resource management tools with the metric as a reference for job allocation and scheduling.

\subsection{Possible Approaches and Challenges}
\label{problemofexisting}
\textbf{Approach 1: Simulation-based Prediction. }
As we mentioned in Section \ref{relatedwork}, one straightforward solution is to estimate the possible result under the current noise parameters using the quantum circuit simulation, which is adopted in the quantum resource management in \cite{ravi2021adaptive}.
We evaluate its efficacy to predict performance bounds as follows.
With the noise model built based on the noise information from a specific quantum backend, the noisy circuit simulations can be performed multiple times to collect the results.
Then, the corresponding $P_{up}(C,N)$ and $P_{low}(C,N)$ can be estimated by taking the maximum and minimum values among them. 

However, to precisely capture a performance bound,
hundreds of runs are needed. As shown in Figure \ref{fig:mot1}(c), the range of the possible values will not converge until sufficient runs, saying 400 for the VQE circuit.
By determining the number of runs, we can obtain a performance bound to serve as a reference for the circuit running result, as shown in Figure \ref{fig:mot1}(d).
This indeed can provide tight bounds with a high $BCR$, satisfying the objectives $1)$ and $2)$ in Definition \ref{def:prob}.

\textbf{Challenge. }
It seems that the noisy simulation can provide a good reference by the identified performance bound; however, it is too costly, which will make the obtained performance out of date.
Specifically, the noise changes frequently, and it might have already reached a new level; however, it's still waiting for the execution results from the local quantum circuit simulator. 
    
\begin{figure*}[t]
    \centering
    \includegraphics[width=1\textwidth]{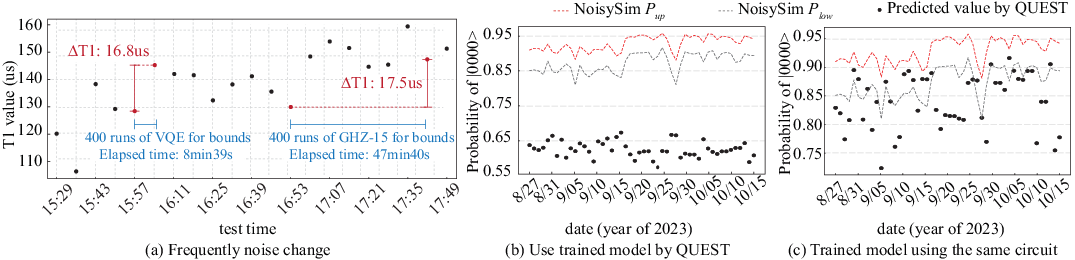}    
    \caption{(a) Frequently noise change makes prediction useless. (b-c) Predicted Results of Existing ML approach - QUEST using VQE-4 circuit, the predicted bounds using noisy simulation serve as the golden result.}
    \label{fig:fig2}
    \vspace{-10pt}
\end{figure*}

We perform noise profiling on IBM \verb|ibmq_kolkata| quantum computer with an interval of 7 minutes, and the results reported in Figure \ref{fig:fig2}(a) show that T1 values change frequently.
This can easily lead to the predicted bounds being useless.
As an example in Figure \ref{fig:fig2}, it needs almost 9 minutes for 400 runs of a 4-qubit VQE circuit, during which $\Delta T1 $ has changed for 16.8 $\mu s$.
Another example is to prepare a GHZ state with 15 qubits: it requires about 48 minutes, during which T1 changes from 130 $\mu s$ to nearly 150 $\mu s$.
What's more, if the circuit qubits exceed the capacity of the quantum simulator, a single simulation will even not yield a result.

\begin{table}[t]
\centering
\caption{Comparison of simulation and theoretical bounds}
\label{tab:motivation3}
\renewcommand{\arraystretch}{1.2}
\tabcolsep 2.8 pt
\small
\begin{tabular}{|c|cc|cccc|}
\hline
\multirow{2}{*}{Circuit} & \multicolumn{2}{c|}{Noisy Sim. (NS) \cite{ravi2021adaptive}} & \multicolumn{4}{c|}{nisqReliability \cite{dasgupta2023reliability}} \\ \cline{2-7} 
 & \multicolumn{1}{c|}{Time} & Range & \multicolumn{1}{c|}{Time} & \multicolumn{1}{c|}{vs. NS} & \multicolumn{1}{c|}{Range} & vs. NS \\ \hline
\multicolumn{1}{|c|}{VQE-4} & \multicolumn{1}{c|}{8 min 39s} & 0.121 & \multicolumn{1}{c|}{0.072s} & \multicolumn{1}{c|}{7,150$\times$} & \multicolumn{1}{c|}{3.194} & 26$\times$ \\ \hline
\multicolumn{1}{|c|}{GHZ-15} & \multicolumn{1}{c|}{47 min 40s} & 0.184 & \multicolumn{1}{c|}{0.255s} & \multicolumn{1}{c|}{11,215$\times$} & \multicolumn{1}{c|}{1.029} & 5.6$\times$ \\ \hline
\end{tabular}
\vspace{-10pt}
\end{table}


\textbf{Approach 2: Analytical-based Prediction.}
The above observation suggests a demand for a more effective predictor 
to reduce the risk of being misled by outdated noise information.
One possible solution to address the efficiency issue is to theoretically calculate performance bounds.
Recent analytical approach nisqReliability\cite{dasgupta2023reliability} can be used to solve the problem.
The idea is to build joint distributions of different noise parameters for the given quantum backend at a past time  $t_1$ and at the run time $t_2$.
Then, the theoretical performance bounds of the circuit can be estimated by comparing the Hellinger difference between the noise distributions at $t_1$ and $t_2$.

\textbf{Challenge.}
Table \ref{tab:motivation3} demonstrates the comparison between the noisy-simulation method and nisqReliability method on the two circuits in terms of performance bound range and execution time. 
Although nisqReliability can establish a boundary range in seconds, which is used to process the mathematical calculations implemented in Python, these theoretical bounds are overly broad.
Taking noisy simulation as the baseline, we observe that the execution time of nisqReliability is much lower, achieving over 7,000$\times$ speedup. However, the ranges of predicted bounds are $26\times$ and $5\times$ larger than the baseline. However, such a wide-range bound is usually not practical.

\textbf{Approach 3: ML-based Prediction.}
As quantum cloud access became available since 2016 \cite{qcloudlist}, the quantum providers have already accumulated enough quantum noise data to be leveraged; for example, IBM provides historical noise data via \verb|IBMQBackend| properties \cite{ibmnoisedata}.
On the other hand, with the rapid development of machine learning in recent years, researchers have investigated how to use a data-driven approach to do performance prediction in HPC systems \cite{yokelson2023hpc}.

It motivates us to rethink --- instead of using theoretical calculation methods, can we use the machine learning method based on the historical noise and corresponding performance traces to resolve the $QuBound_p$ problem? A recent work, namely QUEST\cite{wang2022quest}, predicts the performance of quantum circuits using the graph transformer. It encodes the quantum circuit structure and the noise information as the input and then uses the noisy simulation result as the label.

\textbf{Challenge.}
The ML-based approach can be promising in achieving both accuracy and efficiency; however, some challenges still exist, as described below. Here, the existing method QUEST is used for demonstration, whose results are shown in Figure \ref{fig:fig2}(b-c). We use the predicted bounds from noisy simulation as the golden result, as $P_{up}$ and $P_{low}$ in Figure \ref{fig:fig2}. Two models are trained here, the one shown in Figure \ref{fig:fig2}(b) is the trained model from QUEST that contains 2750 random circuits; while the one shown in Figure \ref{fig:fig2}(c) is the trained model that only uses one specific circuit. The challenges we can learn from the results are:
(1) A single ML model struggles to generalize across all circuits and noise conditions. This is evident in Figure \ref{fig:fig2}(b), where prediction accuracy is low and all dots fall below the lower bound. When trained on a specific circuit, performance improves (Figure \ref{fig:fig2}(c)), but many predictions (black dots) still lie outside the bounds.
(2) Conflicting labels hurt model performance, which could explain the poor results in Figure \ref{fig:fig2}(c). Due to the probabilistic nature of quantum computation, the same circuit under the same noise can yield different outcomes across runs. This results in inconsistent labels for the same input in the built training dataset, which can confuse the learning algorithm and degrade accuracy.
(3) The QUEST method predicts a single value rather than a bound. Although some QUEST predictions fall within the performance bounds (Figure \ref{fig:fig2}(c)), they may still deviate from the actual value that is also within the bounds.



\begin{table}[t]
\centering
\caption{{Comparison of Machine Learning Based Methods}}
\label{tab:newtablemotivation}
\renewcommand{\arraystretch}{1.2}
\tabcolsep 2 pt
\small
\begin{tabular}{|c|cc|cc|cc|}
\hline
\multirow{2}{*}{Circuit} & \multicolumn{2}{c|}{Linear Reg.  (LR) \cite{senapati2024pqml}}  & \multicolumn{2}{c|}{QUEST \cite{wang2022quest}}              & \multicolumn{2}{c|}{QuBound}            \\ \cline{2-7} 
                         & \multicolumn{1}{c|}{Time}      & Error  & \multicolumn{1}{c|}{Time}      & Error  & \multicolumn{1}{c|}{Time}      & Error  \\ \hline
VQE-4                    & \multicolumn{1}{c|}{2.13e-5s} & 0.0437 & \multicolumn{1}{c|}{2.03e-5s} & 0.5109 & \multicolumn{1}{c|}{2.71e-5s} & 0.0183 \\ \hline
GHZ-15                   & \multicolumn{1}{c|}{2.01e-5s} & 0.0292 & \multicolumn{1}{c|}{2.03e-5s} & 0.0401 & \multicolumn{1}{c|}{2.00e-5s} & 0.0238 \\ \hline
\end{tabular}
\vspace{-10pt}
\end{table}

\begin{figure*}[t]
    \centering
    \includegraphics[width=\textwidth]{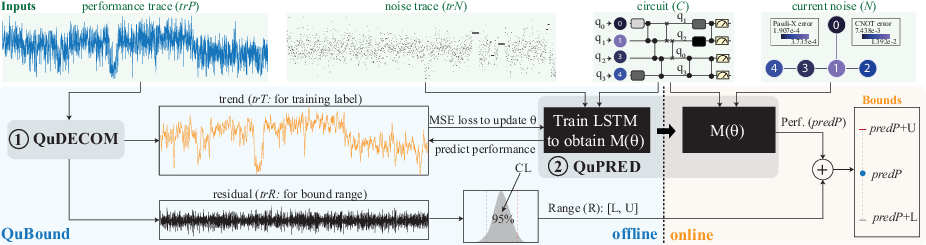}    
    \caption{Overview of the proposed framework \textit{QuBound}, which comprises two components: \textit{QuDECOM}, responsible for decomposing noise-affected results into discernible patterns, and \textit{QuPRED}, dedicated to predicting performance bounds for quantum circuits.}
    \label{fig:overview}
    \vspace{-10pt}
\end{figure*}

\jy{To further explore latency–accuracy tradeoffs, we also evaluate a Linear Regression (LR) method \cite{senapati2024pqml}.
This model maps circuit–noise features to a single predicted performance value without modeling temporal dependencies.
As shown in Table~\ref{tab:newtablemotivation}, LR achieves lower latency (2.01 – 2.13e-5 seconds) than noisy simulation (Table~\ref{tab:motivation3}, 8 – 47 minutes). QUEST and our proposed method \textit{QuBound} also show a similar speedup.}
\jy{However, both LR and QUEST show limited accuracy, with errors ranging from 0.0437 to 0.5109 for circuit VQE-4, significantly higher than those of our model discussed later. 
These results indicate that while ML-based methods can significantly reduce latency compared to simulation, model expressiveness plays a critical role in ensuring prediction reliability.
}

\textbf{The Ideal Approach: a novel workflow for low latency, practical and accurate bounds simultaneously.}

When users need to predict the circuit outcome at a given time, the predictor should be able to give a quick and accurate response. The latency of the predictor should be within an accepted range to protect it from delivering invalid results due to changes in noise information. Also, the range of the predicted bound on the given quantum circuit should be practical and meet users' specifications for different situations.
In the next section, we pilot a novel data-driven workflow to predict quantum performance bounds.

\section{Innovation: Data-driven Q\MakeLowercase{u}B\MakeLowercase{ound}}
\label{section5}
\subsection{System Overview}

\jy{To solve the performance bounding problem defined as $QuBound_p$ in problem definition \ref{def:prob}, we propose a two-stage learning framework called \textit{QuBound}.}
Figure \ref{fig:overview} illustrates the overview of the framework. 
It consists of two components: \textit{QuDECOM} decomposes the noise-influenced results into discernible patterns, while \textit{QuPRED} focuses on predicting the performance bounds of quantum circuits.

\textbf{Offline Phase:}
In the offline phase, three types of data are fed: the performance trace $trP$, the noise trace $trN$, and the quantum circuit $C$. Initially, \textit{QuDECOM} processes the $trP$ to extract two key components: the trend $trT$, which is utilized as the training label, and the residual trace $trR$ used to determine the range of the predicted bounds. Subsequently, \textit{QuPRED} will encode $C$ and $trN$ for the model's input features. This model learns the trend through the minimization of Mean Squared Error (MSE) loss, iteratively refining the model's parameters. Meanwhile, the residual is combined with a selected Confidence Level $CL$ to establish the range for the predicted performance bounds $[L, U]$.

\textbf{Online Phase:}
During the online session, the quantum circuit $C$ and current noise information $N$ are passed into the trained model. This model outputs the predicted performance $predP$. The final prediction bounds $P_{up}(C,N)$ and $P_{low}(C,N)$ are then determined by adding the calculated range $[L, U]$ together, as shown in Figure \ref{fig:overview} right-bottom.
\subsection{Performance Decomposition (QuDECOM)}\label{sec:decomp}



\subsubsection{Multi-Source of Quantum Noise} 
As discussed in Section \ref{problemofexisting}, due to the probabilistic nature of quantum computing, the sampling variation is mixed with the inherent device noise. 


\begin{figure}[t]
    \centering
    \includegraphics[width=0.45\textwidth]{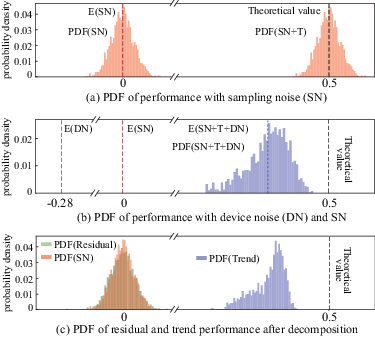}
    \caption{Illustration of performance composition, including theoretical performance (T), performance by sampling noise (SN), and by device noise (DN). }
    \label{fig:decomposition}
    \vspace{-5pt}
\end{figure}

\textbf{Sampling Noise (SN):} The first noise type we consider is sampling noise (SN), which is caused by incomplete measurements as discussed in Section \ref{sec:quantumperformance}. In an ideal simulation scenario (see Figure \ref{fig:decomposition}(a)), the distribution of 2000 simulations for quantum state $|000\rangle$ centers around $0.5$, which is the theoretical value (denoted as $T$) for a GHZ-3 circuit. To isolate SN, we subtract this theoretical value $T$ from the probability distribution function (PDF) of the combined noise and theoretical value $PDF(SN+T)$, resulting in a distribution $PDF(SN)$ centered around zero, as shown by the pale red shape in the left side of Figure \ref{fig:decomposition}(a).

\textbf{Device Noise (DN):} The next source of noise is device noise (DN), originating from the actual quantum device. Another set of 2000 simulations incorporating a noise model gives us the distribution $PDF(SN+T+DN)$, represented in blue in Figure \ref{fig:decomposition}(b). The red dotted line indicates $E(SN)$ derived from ideal circuit simulations. We calculate the expected value of DN as $E(DN) = E(SN+T+DN) - E(SN)-T$. It is observed that $E(DN)$ shifts to the left of $E(SN)$, indicating how DN causes deviations from the theoretical value $T=0.5$.


\begin{algorithm}[t]
\caption{Performance decomposition (QuDECOM)}
\label{alg: alg1}
\small
\begin{algorithmic}[1]
\STATE \textbf{Input:} Historical performance trace $trP$ and the selected confidence level $CL$
\STATE \textbf{Output:} Trend trace $trT$, Bound range [$L$, $U$]

\STATE \textbf{// Outlier Removal:}
\STATE Compute first quartile $Q1$ and third quartile $Q3$ of $trP$
\STATE Compute interquartile range $IQR = Q3 - Q1$
\STATE Remove points from $trP$ outside $[Q1 - 1.5 \cdot IQR, Q3 + 1.5 \cdot IQR]$

\STATE \textbf{// Trend and Residual Extraction:}
\STATE Apply moving average with decomposition period of $dp$ to $trP$ for smoothing data to obtain the trend trace $trT$
\STATE Subtract $trT$ from $trP$ to obtain residual trace $trR$

\STATE \textbf{// Bounds Range Calculation:}
\STATE Calculate the Confidence Interval (CI) for $trR$ using the given $CL$.
\STATE Set the endpoints of $CI$ as $L$ and $U$.

\RETURN $trT$ as training label and [$L$, $U$] as bound range
\end{algorithmic}
\end{algorithm}

\subsubsection{Decomposition Steps}
Algorithm \ref{alg: alg1} shows the details of \textit{QuDECOM}.
We first remove the outliers (lines 3-6). These outliers come from the days when some qubits are broken with an infinite level of noise.
We employ the Interquartile Range (IQR) \cite{NIST_Statistical_Methods} method to identify and remove these outliers.

\jy{Algorithm \ref{alg: alg1}, lines 7-9 show the use} of ``seasonal decomposition'' method to extract the trend and residual components from the performance data. 
We use an additive model since the overall performance is the combination of theoretical performance, performance variation by device noise, and performance variation by sampling noise.
Then, three steps are conducted for decomposition. First, the trend performance is extracted by smoothing the data, using a moving average to the data. To be more specific, the trend performance is extracted using seasonal decomposition with the additive model, implemented via the \texttt{seasonal\_decompose()} function with a parameter of decomposition period $dp$.
Trend refers to the device noise $T+DN$.
Secondly, 
the residual component is the performance after removing the trend component, which is the performance affected by the sampling noise $SN$.
Through this decomposition process, we separate the data into components that align with different noise sources.
This can be observed in Figure \ref{fig:decomposition}(c), where the distribution of residual almost overlaps with the distribution of SN obtained by ideal simulations.


After performance decomposition, the trend component is tied to device-specific noises (such as T1/T2, gate error, etc.); therefore, it best fits as the training labels for the models to predict the performance. On the other hand, the residual component is linked to sampling noise. Thus, it is a suitable choice for generating bound ranges.


\begin{figure}[t]
    \centering
    \includegraphics[width=0.48\textwidth]{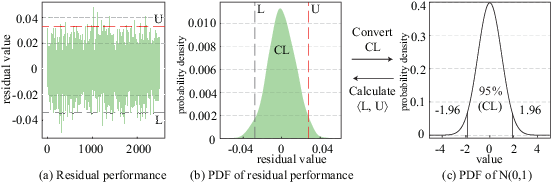}
    \caption{Deriving bound range from the residual performance component}
    \label{fig:bound}
    \vspace{-10pt}
\end{figure}

\subsubsection{Bound Range}
Now, we will show how to use residual performance to get the bound range, shown in Algorithm \ref{alg: alg1} lines 10-12.
The problem is formulated as: Given a required Confidence Level $CL$, the task is to find a range (lower bound $L$ and upper bound $U$) such that the expected residual performance can be in this range with the $CL$.
It involves 4 steps.
First, we obtain the distribution of residual data (with $n$ samples), which exhibits to follow a norm distribution ($N(\mu, \sigma^2)$) as shown in Figure \ref{fig:bound}(b).
Second, we transform to the standardized normal distribution shown in Figure \ref{fig:bound}(c): $\eta = \frac{\bar{X} - E(X)}{\sqrt{\frac{D(X)}{n}}}$ that follows $N(0, 1)$.
Third, 
we obtain the z-score, Z($CL$) by checking the confidence table (e.g., 1.96 for $CL$=0.95); therefore $P( |\eta| \leq 1.96) = 0.95$. 
Last, we can calculate the range of residual performance, which induces $L=\bar{X} - Z(CL) \times \sqrt{\frac{D(X)}{n}}$, and $U=\bar{X} + Z(CL) \times \sqrt{\frac{D(X)}{n}}$.

\subsection{ML-based Predictor (QuPRED)}

\begin{figure}[t]
    \centering
    \includegraphics[width=0.48\textwidth]{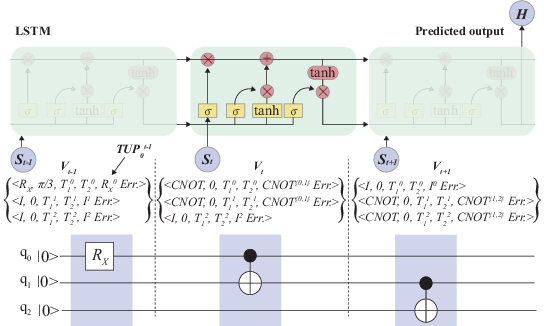}
    \caption{Illustration of encoding quantum circuit for LSTM model.}
    \label{fig:predictor}
    \vspace{-10pt}
\end{figure}

The overview of \textit{QuPRED} is shown in Figure \ref{fig:predictor}, which is composed of an encoder and an LSTM-based learning model.

\textbf{Quantum Circuit and Noise Encoding:}
Our data encoding strategy aims to represent the quantum circuit and noise information in a format that effectively captures the sequential nature of quantum operations and the corresponding noise characteristics.
The $i^{th}$ stage ($S_i$) of the quantum circuit is transformed into a structured vector ($V_i$), encapsulating both the quantum gates and the noise parameters, as shown in Figure \ref{fig:predictor}.
For example, in $S_{t-1}$, we have $V_{t-1}=\{TUP^{t-1}_0,TUP^{t-1}_1,TUP^{t-1}_2\}$, where $TUP_j^{t-1}$ is a 5-tuple corresponding to qubit $q_j$.
For qubit $q_0$, there is one $R_x$ gate, and therefore, we have $TUP_0^{t-1}=\langle R_x,\frac{\pi}{3},T_1^0,T_2^0,Err(R_x^0)\rangle$, where $R_x$ is the gate type, 
\(\frac{\pi}{3}\) is the gate parameter, $T_1^0$ and $T_2^0$ are the relaxation time and dephasing time of $q_0$, and $Err(R_x^0)$ is $R_x$'s error rate on $q_0$.
For qubits $q_1$ and $q_2$, since there are no operation gates, the Identity gate (I Gate) will be applied.
Kindly note that, for a 2-qubit gate, such as the CNOT gate in stage $t$, all elements in $TUP^{t}_0$ and $TUP^{t}_1$ are the same, except different $T1$ and $T2$, since they are two different qubits.
For any given \jy{quantum circuit $C$ with depth $m$} and a given noise $N$, the circuit $C$ can be encoded as a list of structured vectors, denoted as $L_{C,N}=\{V_0,V_1,\cdots,V_{m-1}\}$.
This encoding method is designed to provide both gate and noise information at each stage of the quantum circuit.

By taking the parameters of rotation gates as one of the inputs, the proposed encoder can have more accurate prediction over existing encoding \cite{wang2022quest}; in particular for parameterized quantum circuits. 
By increasing the dimension of the input features, the encoder can be easily extended to include more hyperparameters (such as the gate time, etc.).


\textbf{LSTM Processing of Encoded Data:}
The encoded stages of the quantum circuit are fed sequentially into the LSTM network, where each stage corresponds to a timestamp ($S_{t-1}$, $S_{t}$, $S_{t+1}$) in the LSTM.
By feeding this data through the LSTM, we enable the network to understand not only the individual characteristics of each stage but also the dependencies and interactions between them.
The output from the LSTM, derived from the last time step $H$, serves as the prediction for the quantum circuit’s performance.

\begin{algorithm}[t]
\caption{Quantum Performance Prediction (QuPRED)}
\label{alg:alg2}
\small
\begin{algorithmic}[1]
\STATE \textbf{Input:} Quantum circuit $C$, current noise $N$, noise trace $trN$, corresponding trend data (labels) $trT$
\STATE \textbf{Output:} Predicted performance $predP$

\STATE \textbf{// Prepare Dataset (Output: A training dataset TD)}
\STATE Create an empty training dataset $TD$
\FOR{$\forall (N_i,T_i)\in (trN,trT)$}
    \STATE Add $(L_{C,N_i},T_i)$ into $TD$, where $L_{C,N_i}$ is the circuit and noise encoding.
\ENDFOR


\STATE \textbf{// Model Training (Output: Trained LSTM model M)}
\WHILE{not convergence and preset epoch not reached}
        \STATE Train LSTM model on $TD$ for one batch $B$ to update model weights
\ENDWHILE

\STATE \textbf{// Inference}
\STATE Create circuit and noise encoding $L_{C,N}$
\STATE Predicted performance $predP=M(L_{C,N})$
\RETURN $predP$
\end{algorithmic}
\end{algorithm}

\textbf{QuPRED Algorithm:}
Based on the above understandings, Algorithm \ref{alg:alg2} provides the detailed implementations of \textit{QuPRED}.
It contains three tasks: (1) Lines 3-7, dataset creation based on the historical noise trace $trN$ and performance trend $trT$ decomposed from performance trace $trP$ (see Algorithm \ref{alg: alg1}); (2) Lines 8-11, LSTM model is trained on dataset $TD$; (3) Lines 12-15, inference of LSTM model for the given circuit $C$ and noise $N$.

\textbf{Rationale for the Encoding and LSTM Approach:}
\jy{
In our task, the objective is to predict the performance of a quantum circuit, which is influenced by both the gate operations and the noise conditions during execution. Importantly, these two factors—gate operation and noise—are not independent, but tightly coupled in a sequential manner. The impact of noise on each operation accumulates through the circuit, meaning that earlier noise-gate interactions can affect later operations.
}

\jy{
To model this behavior, we represent each quantum circuit as a sequence of gate operations, where each gate is encoded as a 5-dimensional vector: (1) gate type, (2) gate parameter (e.g., rotation angle), (3) T1 time, (4) T2 time, and (5) gate error rate. 
A full circuit with 10 stages and 3 qubits thus forms a structured 30×5 sequential input array.}
\jy{Given this structure, we choose LSTM as our prediction model due to its inherent ability to capture long-range dependencies across sequences. This aligns with the progressive and accumulative nature of noisy quantum computation, where gate outcomes are context-dependent. By processing the sequence of gates in order, LSTM is capable of learning the temporal correlations introduced by both circuit design and hardware noise.
This choice enables more accurate modeling of circuit behavior under realistic noise, as later shown in our evaluation results.
}


\section{Experimental Results}
\label{section6}
\subsection{Experimental Setup}

\begin{table*}[t]
\centering
\small
\caption{The Comparison between \textit{QuBound} and Existing Method}
\label{tab:comparewithQUEST}
\tabcolsep 3 pt
\renewcommand{\arraystretch}{1.2}
\footnotesize
\begin{tabular}{|c|ccc|ccc|ccc|ccc|ccccc|}
\hline
Circuit & \multicolumn{3}{c|}{NoisySim (baseline)} & \multicolumn{3}{c|}{QUEST} & \multicolumn{3}{c|}{Linear Regression} & \multicolumn{3}{c|}{QuPred} & \multicolumn{5}{c|}{QuBound} \\ \cline{2-18} 
\multicolumn{1}{|l|}{} & \multicolumn{1}{c|}{LB} & \multicolumn{1}{c|}{UB} & Value & \multicolumn{1}{c|}{Value} & \multicolumn{1}{c|}{Diff. 1} & Diff. 2 & \multicolumn{1}{c|}{Value} & \multicolumn{1}{c|}{Diff. 1} & Diff. 2 & \multicolumn{1}{c|}{Value} & \multicolumn{1}{c|}{Diff. 1} & Diff. 2 & \multicolumn{1}{c|}{LB} & \multicolumn{1}{c|}{UB} & \multicolumn{1}{c|}{Value} & \multicolumn{1}{c|}{Diff. 1} & Diff. 2 \\ \hline
GHZ-3 & \multicolumn{1}{c|}{0.8955} & \multicolumn{1}{c|}{0.9345} & 0.9162 & \multicolumn{1}{c|}{0.9546} & \multicolumn{1}{c|}{0.0201} & 0.0384 & \multicolumn{1}{c|}{0.8981} & \multicolumn{1}{c|}{0} & 0.0181 & \multicolumn{1}{c|}{0.9204} & \multicolumn{1}{c|}{0} & 0.0042 & \multicolumn{1}{c|}{0.8980} & \multicolumn{1}{c|}{0.9375} & \multicolumn{1}{c|}{0.9178} & \multicolumn{1}{c|}{0} & 0.0016 \\ \hline
RB-3 & \multicolumn{1}{c|}{0.6805} & \multicolumn{1}{c|}{0.7855} & 0.7387 & \multicolumn{1}{c|}{0.2110} & \multicolumn{1}{c|}{0.4695} & 0.5277 & \multicolumn{1}{c|}{0.7548} & \multicolumn{1}{c|}{0} & 0.0161 & \multicolumn{1}{c|}{0.7294} & \multicolumn{1}{c|}{0} & 0.0093 & \multicolumn{1}{c|}{0.6645} & \multicolumn{1}{c|}{0.8149} & \multicolumn{1}{c|}{0.7397} & \multicolumn{1}{c|}{0} & 0.0010 \\ \hline
GHZ-4 & \multicolumn{1}{c|}{0.8595} & \multicolumn{1}{c|}{0.9025} & 0.8837 & \multicolumn{1}{c|}{0.6948} & \multicolumn{1}{c|}{0.1647} & 0.1889 & \multicolumn{1}{c|}{0.8111} & \multicolumn{1}{c|}{0.0484} & 0.0727 & \multicolumn{1}{c|}{0.8966} & \multicolumn{1}{c|}{0} & 0.0129 & \multicolumn{1}{c|}{0.8623} & \multicolumn{1}{c|}{0.9093} & \multicolumn{1}{c|}{0.8858} & \multicolumn{1}{c|}{0} & 0.0021 \\ \hline
HS-4 & \multicolumn{1}{c|}{0.7800} & \multicolumn{1}{c|}{0.8305} & 0.8049 & \multicolumn{1}{c|}{0.3665} & \multicolumn{1}{c|}{0.4135} & 0.4384 & \multicolumn{1}{c|}{0.7798} & \multicolumn{1}{c|}{0.0002} & 0.0251 & \multicolumn{1}{c|}{0.8236} & \multicolumn{1}{c|}{0} & 0.0187 & \multicolumn{1}{c|}{0.7830} & \multicolumn{1}{c|}{0.8521} & \multicolumn{1}{c|}{0.8175} & \multicolumn{1}{c|}{0} & 0.0126 \\ \hline
VQE-4 & \multicolumn{1}{c|}{0.8510} & \multicolumn{1}{c|}{0.894} & 0.8698 & \multicolumn{1}{c|}{0.4049} & \multicolumn{1}{c|}{0.4461} & 0.4649 & \multicolumn{1}{c|}{0.8261} & \multicolumn{1}{c|}{0.0249} & 0.0437 & \multicolumn{1}{c|}{0.9117} & \multicolumn{1}{c|}{0.0177} & 0.0420 & \multicolumn{1}{c|}{0.8651} & \multicolumn{1}{c|}{0.9111} & \multicolumn{1}{c|}{0.8881} & \multicolumn{1}{c|}{0} & 0.0183 \\ \hline
QAOA-4 & \multicolumn{1}{c|}{0.5170} & \multicolumn{1}{c|}{0.653} & 0.5892 & \multicolumn{1}{c|}{0.1442} & \multicolumn{1}{c|}{0.3728} & 0.4449 & \multicolumn{1}{c|}{0.6341} & \multicolumn{1}{c|}{0} & 0.0449 & \multicolumn{1}{c|}{0.5517} & \multicolumn{1}{c|}{0} & 0.0375 & \multicolumn{1}{c|}{0.5241} & \multicolumn{1}{c|}{0.6697} & \multicolumn{1}{c|}{0.5969} & \multicolumn{1}{c|}{0} & 0.0077 \\ \hline
\end{tabular}
\vspace{-10pt}
\end{table*}
\textbf{Benchmarks:}
We employ widely used benchmarks for the evaluation, including the Greenberger–Horne–Zeilinger (GHZ) \cite{greenberger1989going} circuit used for quantum state preparation, the Randomized Benchmarking (RB) circuit from Qiskit Experiments \cite{kanazawa2023qiskit}, and the quantum circuits generated from SupermarQ \cite{tomesh2022supermarq} including Hamiltonian Simulation (HS), Vanilla QAOA (QAOA), and Variational Quantum Eigensolver (VQE).
Different numbers of qubits are applied, denoted as Benchmark-Qubit, saying ``GHZ-3'' indicates the GHZ circuit with three qubits. Notice that to use the existing ML-based approach - QUEST, the above circuits will be appended with their reverse circuits during the experiments, compared with QUEST.

\textbf{Metrics:}
For comparison, we use three metrics to evaluate the practicality of the methods, including \textit{i)} bound range (denoted as ``Range'') between the predicted lower bound ($P_{low}$) and upper bound ($P_{up}$), \textit{ii)} Bound-compliance rate (denoted as ``$BCR$''), and \textit{iii)} elapsed time in seconds (denoted as ``Time'') refers to the end-to-end execution time.

To compare with QUEST, we define two metrics: Difference 1 (denoted as `Diff. 1') and Difference 2 (denoted as `Diff. 2'). Diff. 1 evaluates whether a predicted value \( V \) is within the range defined by \( LB \) and \( UB \). It is calculated as follows: if \( LB \leq V \leq UB \), then \(\text{Diff. 1} = 0\); otherwise, \(\text{Diff. 1} = \min(|V - LB|, |V - UB|)\). This metric outputs zero if the value is within the range, otherwise, it returns the smallest deviation from the boundaries. Diff. 2 calculates the absolute difference between the predicted value and the baseline value \( V \) and \( V_{baseline} \), defined as \(\text{Diff. 2} = |V - V_{baseline}|\).

\textbf{\textit{QuBound} Configurations:}
\jy{
In \textit{QuDECOM}, the decomposition period is set to 7 based on extensive experiments. \textit{QuPRED} uses a unidirectional LSTM with one hidden layer (size 64). The input sequence for each circuit is constructed by flattening gate-wise feature vectors (each of 5 dimensions) across all qubits and timesteps, resulting in an input shape of $(\text{batch}, \text{stages}, \text{num\_qubits} \times 5)$. The LSTM output is passed through a sequence of fully connected layers with dimensions 128 → 64 → ${\text{number of quantum states}}$, with dropout (0.2) applied between layers. We train the model using the Adam optimizer with a learning rate of 0.008, batch size of 64, and up to 200 epochs using the MSLE loss.
}

\textbf{Approaches for Comparison:}
We involve four performance prediction methods in the experiments, including \textit{i)} the noisy simulation method \cite{ravi2021adaptive} as the baseline, denoted as ``NoisySim'', \textit{ii)} the analytical approach proposed in \cite{dasgupta2023reliability}, denoted as ``nisqReliability'', \textit{iii)} the existing ML-based approach proposed in \cite{wang2022quest}, denoted as ``QUEST'', and \textit{(iv)} our proposed \textit{QuBound}.
For \textit{QuBound}, as it provides flexibility in specifying confidence level (CL), we denote ``\textit{QuBound}-X'' to indicate the $CL=X\%$, say ``\textit{QuBound}-90'' stands for the confidence level set to be 90\% in the bound prediction.

\textbf{Dataset and model:} We collect the noise historical data using the IBM Qiskit, leveraging noisy simulation to generate the performance trace. 
Note that our method is not limited to this; the performance trace can be replaced by data obtained from actual quantum computers.
The embedding of historical noise and target circuit is the dataset input, with the trend performance decomposed by \textit{QuDECOM} as labels.
The dataset contains 2800 samples, collected over 700 days (4 noise samples/day), and is split into 2600 for training and 200 for testing.
We also generate the results to be predicted by running benchmarks on real quantum devices, the \verb|ibm_nairobi| and \verb|ibmq_mumbai|. The trained model from QUEST contains 2750 random circuits in the training dataset\jy{\footnote{Datasets available at
\url{github.com/jinyang001/Datasets-for-QuBound}}}.

\subsection{Results Using Probability-based Performance}
\label{section6b}



\subsubsection{Comparison with Existing Machine Learning Method}
In this section, we compare \textit{QuBound} with the existing ML-based approach, QUEST. The results are shown in Table \ref{tab:comparewithQUEST}. Here we use the noisy simulation results as the baseline. Since QUEST only predicts a single value, we use the mean value of all simulation results as the baseline value, which is shown in the column `Value'. The column `Value' under `QUEST' refers to QUEST's predicted value. Then the columns `\textit{QuPRED}' and `\textit{QuBound}' show the results that only use the \textit{QuPRED} component and the results that use both the \textit{QuPRED} and \textit{QuDECOM} components, separately. Also, in order to compare with QUEST, we only use the predicted value of the \textit{QuBound} instead of a predicted performance bound. \jy{This predicted value corresponds to the central point of the performance bound estimated by \textit{QuBound}, which we denote as \textit{predP} in Fig.~\ref{fig:overview} (bottom-right corner). As shown in the figure, our model outputs not only this single value prediction, but also the associated lower and upper bounds. To ensure a fair comparison with QUEST—which only produces a single scalar value—we use \textit{predP} as the comparison target. In Table~\ref{tab:comparewithQUEST}, this value is reported under the `Value' column for `\textit{QuBound}', consistent with the other methods. Meanwhile, the bounds (`LB' and `UB') are also included in the same table to reflect the complete capability of our approach.
} \jy{In addition to QUEST, we also include another method based on Linear Regression, as applied in prior work \cite{senapati2024pqml}. This method uses the same input features as \textit{QuPRED}—the extracted performance traces from quantum circuit executions—but applies a linear model to predict the expected value. Similar to QUEST, it does not output a bound and only produces a single scalar value.}

As we can see from the table, none of the predicted values from QUEST successfully fall into the baseline bound. While only the predicted result on GHZ-3 is closer to the baseline bounds and value, with a Diff. 1 = 0.0201 and Diff. 2 = 0.0384. \jy{The Linear Regression method, owing to its focus on the target circuit rather than learning a global model across circuits, performs better than QUEST. However, due to its inability to capture sequential gate-noise interactions, it still shows significant deviation from the baseline bounds in most cases. For instance, on GHZ-4 and VQE-4, their predicted values lie well outside the baseline range. This highlights its limited capacity, even when using the same input features as \textit{QuPRED}.} 
Then, for the results only using the \textit{QuPRED} component, almost all the predicted values are within the baseline bound, while only the value on VQE-4 has a very small deviation from the bound. The predicted values by \textit{QuPRED} are much closer to the baseline value compared with the result from QUEST. 
Last, after we use both the two components of \textit{QuBound}, all predicted results fall into the baseline bound. Also, the predicted value is even closer to the baseline value compared with only using \textit{QuPRED}.

We can learn from the results that: First, the importance of circuit-specific when choosing the ML method to predict the performance of quantum circuits, which can be proved by the huge prediction accuracy difference in the comparison between QUEST and \textit{QuPRED}. \jy{Second, although Linear Regression uses the same input data as \textit{QuPRED}, its simple model structure fails to generalize to more complex quantum behavior. This suggests that even with relevant features, the model’s expressive power plays a crucial role in achieving accurate predictions.
Third}, the effectiveness of the \textit{QuDECOM} when helping improve the prediction accuracy.

\begin{figure}[t]
    \centering
    \includegraphics[width=0.48\textwidth]{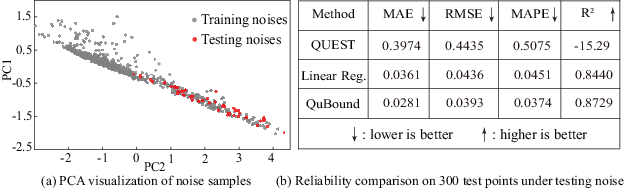}
    \caption{
    \jy{Reliability evaluation of \textit{QuBound} under testing noise.
    }}
    \label{fig:pca-reliability}
    \vspace{-10pt}
\end{figure}
\subsubsection{Reliability Evaluation under Testing Noise Conditions}

\jy{
To evaluate the model's reliability and generalization, we conduct an experiment using only testing noise.}
\jy{
Specifically, we collect 50 testing-day noise traces for the six benchmark circuits, totaling 300 test samples. For each circuit under each day's noise, we compare the prediction accuracy of QUEST, Linear Regression, and \textit{QuBound}. All predictions are compared against the noisy simulation results as ground truth.}

\jy{
In Figure \ref{fig:pca-reliability}(a), we visualize the distribution of training and testing noise samples using PCA, which reduces the 19-dimensional noise vector to 2D for illustration. Red and gray dots represent testing and training noise, respectively. The separation between them highlights the generalization challenge.} 
\jy{
Figure \ref{fig:pca-reliability}(b) shows the prediction accuracy of the three methods using four standard regression metrics (MAE, RMSE, MAPE, and R$^2$). 
\textit{QuBound} outperforms both QUEST and Linear Regression across all metrics. Notably, \textit{QuBound} achieves the lowest MAE of 0.0281 and the highest R$^2$ of 0.8729, while QUEST exhibits poor generalization with an R$^2$ below zero.
}
\jy{
These results show that \textit{QuBound} maintains high prediction reliability under unseen noise.
}

\subsubsection{Scalability Results}

\begin{figure}[t]
    \centering
    \includegraphics[width=0.48\textwidth]{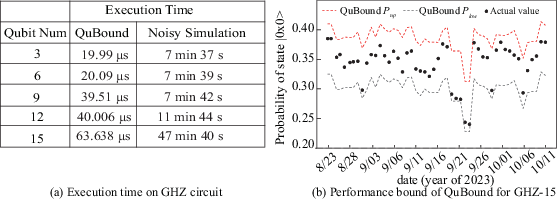}
    \caption{Test scalability of \textit{QuBound} on larger circuits.}
    \label{fig:scalbility}
    \vspace{-10pt}
\end{figure}

Figure \ref{fig:scalbility} reports the scalability of \textit{QuBound}. 
The left table shows the execution time of \textit{QuBound} and noisy simulation as qubit count increases, while the right plot visualizes the predicted bound of \textit{QuBound} on a larger circuit, GHZ-15. The execution time of \textit{QuBound} is slightly increasing but still on a very small scale, from 19.99 $\mu s$ to 63.638 $\mu s$. On the other hand, the noisy simulation method will jump up to over 47 minutes as the number of qubits reaches 15. Overall, the time for \textit{QuBound} is much shorter than the noisy simulation method. For the visualization of performance on the right figure, \textit{QuBound} can predict a practical bound on the circuit with 15 qubits. The result further validates the effectiveness and efficiency of \textit{QuBound} on larger circuits.
We limit the scale to 15 qubits due to the large simulation cost.
However, since \textit{QuBound} can also use real quantum hardware to build performance traces, it remains applicable at much larger scales beyond what simulation permits.

\subsubsection{Different Backends and Noise Type Results}

\begin{table}[t]
\centering
\caption{Performance of QuBound on different backends}
\label{tab:different_backend}
\footnotesize
\tabcolsep 2.5 pt
\renewcommand{\arraystretch}{1.2}
\begin{tabular}{|c|cc|cc|cc|}
\hline
\multicolumn{1}{|l|}{} & \multicolumn{2}{c|}{\textit{QuBound}-90} & \multicolumn{2}{c|}{\textit{QuBound}-95} & \multicolumn{2}{c|}{\textit{QuBound}-99} \\ \hline
Backend & \multicolumn{1}{c|}{$BCR$} & $Range$ & \multicolumn{1}{c|}{$BCR$} & $Range$ & \multicolumn{1}{c|}{$BCR$} & $Range$ \\ \hline
\verb|ibm_hanoi| & \multicolumn{1}{c|}{86.75\%} & 0.01778 & \multicolumn{1}{c|}{93.25\%} & 0.02118 & \multicolumn{1}{c|}{99.25\%} & 0.03340 \\ \hline
\verb|ibmq_kolkata| & \multicolumn{1}{c|}{86.50\%} & 0.01584 & \multicolumn{1}{c|}{93.75\%} & 0.01888 & \multicolumn{1}{c|}{99.75\%} & 0.02977 \\ \hline
\verb|ibmq_mumbai| & \multicolumn{1}{c|}{88.00\%} & 0.01890 & \multicolumn{1}{c|}{92.75\%} & 0.02252 & \multicolumn{1}{c|}{99.00\%} & 0.03551 \\ \hline
\verb|ibm_nairobi| & \multicolumn{1}{c|}{90.00\%} & 0.02131 & \multicolumn{1}{c|}{93.25\%} & 0.02539 & \multicolumn{1}{c|}{99.75\%} & 0.04004 \\ \hline
\end{tabular}
\vspace{-10pt}
\end{table}

\begin{figure*}[t]
    \centering
    \includegraphics[width=1\textwidth]{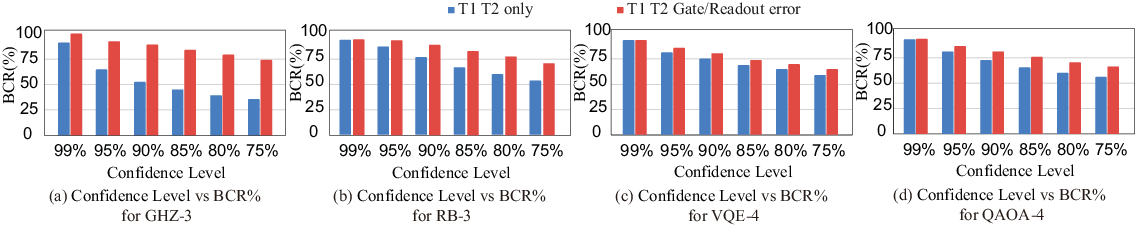}
    \caption{Ablation study on the impact of different types of noise by comparing the performance under two different LSTM encodings: (1) using only T1 and T2, and (2) incorporating T1, T2, gate errors, and readout errors.}
    \label{fig:abstudy1}
    \vspace{-10pt}
\end{figure*}

\begin{figure*}[t]
    \centering
    \includegraphics[width=1\textwidth]{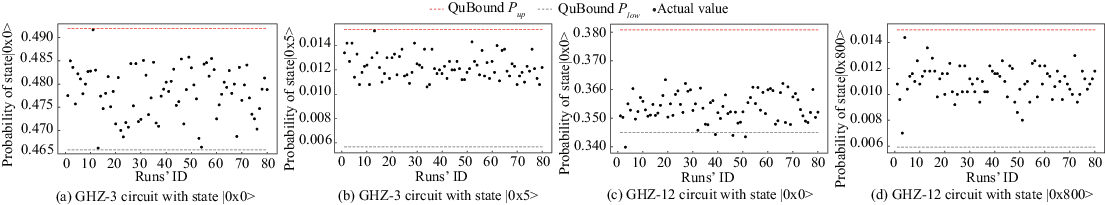}
    \caption{Performance prediction results on real quantum device for GHZ-3 and GHZ-12 circuits (IBM ibmq\_mumbai on March 30, 2024).}
    \label{fig:new_real_result}
    \vspace{-10pt}
\end{figure*}



Table \ref{tab:different_backend} summarizes the performance outcomes of \textit{QuBound} across various backends. We chose three different confidence intervals for the predicted bound range.
The experiments are conducted using 200 samples from the created dataset.
It is evident that \textit{QuBound} maintains consistently high performance across all tested backends, albeit with minor variations in the bound range. 
From the results, we can observe that the $BCR$ value is quite close to the confidence interval; especially for \textit{QuBound}-99, the deviation between $BCR$ and the confidence interval is merely 0.44.
For \textit{QuBound}-90 and \textit{QuBound}-95, the average deviation is 2.19 and 1.75, respectively.
This again shows the effectiveness of the performance decomposition \textit{QuDECOM}, which can work together with \textit{QuPRED} to identify meaningful bound, which is consistent with statistics.

Figure \ref{fig:abstudy1} compares the $BCR$ of \textit{QuBound} under two LSTM encoding schemes: (1) employing only T1 and T2, and (2) incorporating T1, T2, gate errors and readout errors. Both approaches can yield satisfactory results across four circuits when a high confidence level (CL) is chosen. However, a notable divergence is observed as the CL is lowered. Specifically, the approach using only T1 and T2 values experiences a more pronounced decline in $BCR$ compared to the method that also considers gate errors. This is evident in the GHZ-3 circuit, as Figure \ref{fig:abstudy1}(a) depicts. When the CL is reduced below 90\%, the $BCR$ of the T1 and T2 only method falls below 50\%. Results show the impact of noise models on performance, as \textit{QuBound} has the ability to holistically take different error sources into consideration, thus yielding enhanced performance.

\subsubsection{Results on Actual Quantum Computer}
We conduct experiments to further show the scalability of \textit{QuBound} on real quantum processors, and the results are shown in Figure \ref{fig:new_real_result}.
We predict the GHZ-3 circuit on IBM \verb|ibmq_mumbai| on March 30, 2024.
Figures \ref{fig:new_real_result}(a)-(b) show the prediction results for GHZ-3 circuit on two states, $|000\rangle$ and $|101\rangle$.
The circuit runs 80 times, each of which is presented as a point in the figures.
As all circuits are submitted and executed in the same job, the bounds are fixed, as shown in the red and black lines.
We can see that \textit{QuBound} can generate reasonable bounds to cover all actual performance.
We then evaluate the scalability of \textit{QuBound} on a GHZ-12 circuit.
Figures \ref{fig:new_real_result}(c)-(d) demonstrate the results for two states.
As shown in these two figures, there are only 4 runs that exceed the computational bound for state $|0\cdots0\rangle$, while all runs are within the bounds for the other state.
All results above verify the proposed \textit{QuBound} can work for the actual quantum processors.

\subsection{Results Using Observable-based Performance}

\subsubsection{Effectiveness Evaluation}
Table \ref{tab:bigtable} reports the performance comparison of the three prediction methods.
The column `Noisy Result' refers to the performance result from the test dataset, which is obtained by the $Z^{\otimes N}$ observable, where $N$ is the number of qubits.

As we can see from the table, the predicted bounds of all three methods can cover the `Noisy Results' on all circuits. 
However, they vary in the range of bounds and elapsed time to obtain the bounds. 
Then, the baseline method has the narrowest range on all circuits except the GHZ-3 circuit; however, its elapsed time is over 500 seconds. 
Second, the nisqReliability method and the \textit{QuBound} make predictions in a very short time period, which achieves 7,970$\times$ and 942,000$\times$ speedup on average over baseline.
Third, nisqReliability has a very wide bound range on all circuits, which is more than 24$\times$ larger than the baseline on average. In contrast, \textit{QuBound} with a 90\% confidence level can obtain a range of bounds close to the baseline on all circuits. It even obtains a narrower range on circuit GHZ3 (0.152 vs. 0.154).
The above results show that \textit{QuBound} is more efficient than NoisySim and can find more practical bounds than nisqReliability. 

\begin{table*}[t]
\centering
\scriptsize
\caption{The comparison of \textit{QuBound} with confidence level of 90\% to noisy simulation and analytical approach on various circuits}
\label{tab:bigtable}
\tabcolsep 4.9 pt
\renewcommand{\arraystretch}{1.2}

\begin{tabular}{|c|c|cccc|cccccc|cccccc|}
\hline
\multirow{2}{*}{\begin{tabular}[c]{@{}c@{}}Circuit\\ \end{tabular}} & \multirow{2}{*}{\begin{tabular}[c]{@{}c@{}}Noisy\\ Result\end{tabular}} & \multicolumn{4}{c|}{NoisySim (baseline)} & \multicolumn{6}{c|}{nisqReliability} & \multicolumn{6}{c|}{\textit{QuBound}-90} \\ \cline{3-18} 
 &  & \multicolumn{1}{c|}{$P_{low}$} & \multicolumn{1}{c|}{$P_{up}$} & \multicolumn{1}{c|}{Range} & Time & \multicolumn{1}{c|}{$P_{low}$} & \multicolumn{1}{c|}{$P_{up}$} & \multicolumn{1}{c|}{Range} & \multicolumn{1}{c|}{vs.bl} & \multicolumn{1}{c|}{Time} & vs.bl & \multicolumn{1}{c|}{$P_{low}$} & \multicolumn{1}{c|}{$P_{up}$} & \multicolumn{1}{c|}{Range} & \multicolumn{1}{c|}{vs.bl} & \multicolumn{1}{c|}{Time} & vs.bl \\ \hline
GHZ-3 & -0.068 & \multicolumn{1}{c|}{-0.079} & \multicolumn{1}{c|}{0.075} & \multicolumn{1}{c|}{0.154} & 5.11e2 & \multicolumn{1}{c|}{-0.481} & \multicolumn{1}{c|}{0.557} & \multicolumn{1}{c|}{1.039} & \multicolumn{1}{c|}{6.75} & \multicolumn{1}{c|}{5.83e-2} & 8.77e3 & \multicolumn{1}{c|}{-0.069} & \multicolumn{1}{c|}{0.083} & \multicolumn{1}{c|}{0.152} & \multicolumn{1}{c|}{0.99} & \multicolumn{1}{c|}{4.00e-5} & 1.28e7 \\ \hline
RB-3 & 0.774 & \multicolumn{1}{c|}{0.723} & \multicolumn{1}{c|}{0.827} & \multicolumn{1}{c|}{0.104} & 5.33e2 & \multicolumn{1}{c|}{-1.260} & \multicolumn{1}{c|}{2.736} & \multicolumn{1}{c|}{3.996} & \multicolumn{1}{c|}{38.42} & \multicolumn{1}{c|}{5.68e-2} & 9.38e3 & \multicolumn{1}{c|}{0.635} & \multicolumn{1}{c|}{0.857} & \multicolumn{1}{c|}{0.222} & \multicolumn{1}{c|}{2.13} & \multicolumn{1}{c|}{1.40e-4} & 3.81e6 \\ \hline
GHZ-4 & 0.838 & \multicolumn{1}{c|}{0.750} & \multicolumn{1}{c|}{0.842} & \multicolumn{1}{c|}{0.092} & 5.35e2 & \multicolumn{1}{c|}{-1.102} & \multicolumn{1}{c|}{2.796} & \multicolumn{1}{c|}{3.898} & \multicolumn{1}{c|}{42.37} & \multicolumn{1}{c|}{6.99e-2} & 7.66e3 & \multicolumn{1}{c|}{0.754} & \multicolumn{1}{c|}{0.942} & \multicolumn{1}{c|}{0.188} & \multicolumn{1}{c|}{2.04} & \multicolumn{1}{c|}{4.00e-5} & 1.34e7 \\ \hline
HS-4 & 0.208 & \multicolumn{1}{c|}{0.131} & \multicolumn{1}{c|}{0.279} & \multicolumn{1}{c|}{0.148} & 5.34e2 & \multicolumn{1}{c|}{-0.850} & \multicolumn{1}{c|}{1.254} & \multicolumn{1}{c|}{2.103} & \multicolumn{1}{c|}{14.21} & \multicolumn{1}{c|}{7.16e-2} & 7.45e3 & \multicolumn{1}{c|}{0.068} & \multicolumn{1}{c|}{0.335} & \multicolumn{1}{c|}{0.267} & \multicolumn{1}{c|}{1.81} & \multicolumn{1}{c|}{4.00e-5} & 1.33e7 \\ \hline
VQE-4 & 0.460 & \multicolumn{1}{c|}{0.347} & \multicolumn{1}{c|}{0.468} & \multicolumn{1}{c|}{0.121} & 5.19e2 & \multicolumn{1}{c|}{-1.155} & \multicolumn{1}{c|}{2.039} & \multicolumn{1}{c|}{3.194} & \multicolumn{1}{c|}{26.39} & \multicolumn{1}{c|}{7.26e-2} & 7.15e3 & \multicolumn{1}{c|}{0.391} & \multicolumn{1}{c|}{0.592} & \multicolumn{1}{c|}{0.201} & \multicolumn{1}{c|}{1.66} & \multicolumn{1}{c|}{6.00e-5} & 8.66e6 \\ \hline
QAOA-4 & -0.124 & \multicolumn{1}{c|}{-0.220} & \multicolumn{1}{c|}{-0.081} & \multicolumn{1}{c|}{0.139} & 5.32e2 & \multicolumn{1}{c|}{-1.314} & \multicolumn{1}{c|}{0.946} & \multicolumn{1}{c|}{2.261} & \multicolumn{1}{c|}{16.26} & \multicolumn{1}{c|}{7.20e-2} & 7.39e3 & \multicolumn{1}{c|}{-0.315} & \multicolumn{1}{c|}{0.037} & \multicolumn{1}{c|}{0.352} & \multicolumn{1}{c|}{2.53} & \multicolumn{1}{c|}{1.17e-4} & 4.55e6 \\ \hline
\end{tabular}
\vspace{-10pt}
\end{table*}

\begin{figure*}[t]
    \centering
    \includegraphics[width=1\textwidth]{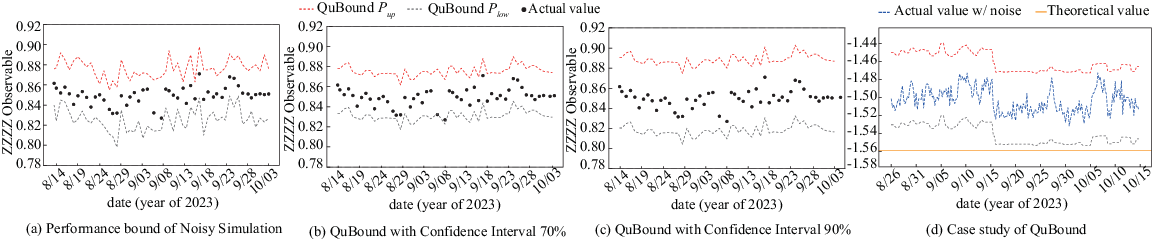}
    \caption{Visualization of results of \textit{QuBound} and NoiseSim over 50 days: (a) NoiseSim; (b) \textit{QuBound}-70; (c) \textit{QuBound}-90; (d) Case study of \textit{QuBound} predicting $H_2$ ground state energy using Variational Quantum Eigensolver, a fundamental quantum chemistry problem.}
    \label{fig:visualization}
    \vspace{-10pt}
\end{figure*}

\subsubsection{Visualization Results}
Figure \ref{fig:visualization} illustrates the visualization results of \textit{QuBound} and NoisySim under the different noises over 50 days. The Y-axis represents the value of $ZZZZ$ (or $Z^{\otimes 4}$) observable on the circuit.
Figure \ref{fig:visualization}(a) shows the bounds obtained by the noisy simulation, while Figures \ref{fig:visualization}(b)-(c) give the results of \textit{QuBound} with the choice of different confidence intervals.
From these figures, we can observe that \textit{QuBound} can capture the noise properties and obtain a practical bound range. 
As shown in Figure \ref{fig:visualization}(b), if we decrease the confidence interval to 70\%, \textit{QuBound} can achieve a bound range that is even narrower than the noisy simulation method while still being able to cover all points.
These results show that both \textit{QuBound} and noisy simulation can predict the circuit outcome under different noises, while \textit{QuBound} can have a flexible control on the range of predicted bounds.

\subsubsection{Prediction on Actual IBM Quantum Machine}

Table \ref{tab:real} summarizes the performance result of \textit{QuBound}'s prediction for a GHZ-3 circuit on an actual quantum device (\verb|ibm_nairobi|), where the confidential level is set as 99\%. 
We record the submission date and the exact running date in the table. Each day, we submitted 80 circuits, and the column `Actual Min' and `Actual Max' are the minimum and maximum from the obtained results. From the table, we can observe that with the predicted bound, the $BCR$ is 100 on most days, indicating all 80 measurement results obtained by the real quantum device fall into the predicted bounds.
Interestingly, two jobs submitted on Oct. 23 and Oct. 24 were postponed to be executed on Oct. 25, and a couple of results obtained by these jobs completed on Oct. 25 are falling outside the bounds. 
This is possible because of the heavy workload for the backend during Oct. 23 - Oct. 24, which led to inaccurate device calibration.
All the above results demonstrate that our proposed \textit{QuBound} can be effectively used in the actual quantum execution environment.



\begin{table}[t]
\centering
\caption{Test result on the real quantum device in Oct., 2023}
\label{tab:real}
\tabcolsep 3.9 pt
\footnotesize
\renewcommand{\arraystretch}{1.2}
\begin{tabular}{|c|c|c|c|c|c|c|}
\hline
\begin{tabular}[c]{@{}c@{}}Submit\\  Date\end{tabular} & \begin{tabular}[c]{@{}c@{}}Runing\\ date\end{tabular} & \begin{tabular}[c]{@{}c@{}}Actual\\ Min\end{tabular} & \begin{tabular}[c]{@{}c@{}}Actual\\  Max\end{tabular} & \begin{tabular}[c]{@{}c@{}}\textit{QuBound}\\ $P_{low}$\end{tabular} & \begin{tabular}[c]{@{}c@{}}\textit{QuBound}\\  $P_{up}$\end{tabular} & BCR (\%) \\ \hline
10/20 & 10/20 & 0.109 & 0.153 & 0.027 & 0.167 & 100 \\ \hline
10/21 & 10/21 & 0.097 & 0.149 & 0.017 & 0.156 & 100 \\ \hline
10/22 & 10/22 & 0.063 & 0.112 & 0.015 & 0.155 & 100 \\ \hline
10/23 & \textbf{10/25} & 0.110 & 0.164 & 0.014 & 0.153 & \textbf{96.25} \\ \hline
10/24 & \textbf{10/25} & 0.1146 & 0.1836 & 0.014 & 0.153 & \textbf{88.75} \\ \hline
10/25 & \textbf{10/25} & 0.1024 & 0.1628 & 0.014 & 0.153 & \textbf{96.25} \\ \hline
10/26 & 10/26 & 0.104 & 0.146 & 0.017 & 0.156 & 100 \\ \hline
10/27 & 10/28 & 0.098 & 0.147 & 0.010 & 0.149 & 100 \\ \hline
10/28 & 10/28 & 0.098 & 0.1464 & 0.010 & 0.149 & 100 \\ \hline
10/29 & 10/29 & 0.111 & 0.158 & 0.011 & 0.151 & 98.75 \\ \hline
10/30 & 11/04 & 0.107 & 0.156 & 0.024 & 0.164 & 100 \\ \hline
\end{tabular}
\vspace{-10pt}
\end{table}

\subsubsection{Case Study on VQE for Ground-state Energy}


This subsection applies the proposed \textit{QuBound} to calculate the ground-state energy of the Hydrogen (H2) molecule using VQE, which is a real-world quantum chemistry application.
Note that there are a couple of reasons that the quantum circuit for VQE needs to be reused: (1) the trained VQE ansatz can be reused as a starting point for exploring excited states, and (2) a trained VQE ansatz can potentially be reused for screening similar molecules.
In these scenarios, understanding the performance bounds of VQE ansatz is of utmost importance.

Figure \ref{fig:visualization}(d) reports the result of a trained 
parameterized VQE circuit (ansatz) to calculate the ground state energy of the H2 molecule, where the 
qubit operator is represented by the following Pauli strings: $(-1.052373245772859 \times I \otimes I) + (0.39793742484318045 \times I \otimes Z) + (-0.39793742484318045 \times Z \otimes I) + (-0.01128010425623538 \times Z \otimes Z) + (0.18093119978423156 \times X \otimes X)$.
To enable \textit{QuBound}, we will perform the bound prediction for each observable (say $X\otimes X$).
Then, we sum up the upper bounds (lower bounds) to obtain the final bounds, which are plotted as red and grey dotted lines.
In the figure, we also show the theoretical value of the VQE as the orange solid line, and the blue dotted line shows the performance of the trained ansatz changes under a noisy execution environment over 50 days.
Obviously, \textit{QuBound} can perfectly cover the blue lines, which indicates that \textit{QuBound} can accurately predict the energy outcome.

\section{Discussion}
\label{section7}


\textbf{Practicability of Solver in Real World.} An efficient and accurate solver to problem $QuBound_p$ can predict the performance bound for a given quantum circuit without historical performance trace $trP$. Such a use case can be applied to different applications, such as circuit compilation and resource management, where the system does not have the historical information of the on-the-fly incoming quantum circuits.

\jy{To be more specific, 
one prominent use case is in multi-backend job scheduling: when multiple backends with different noise profiles are available, system schedulers must decide where to execute each incoming job to maximize fidelity. Our method can rapidly estimate their performance bounds, enabling informed backend selection.
Another use case is circuit compilation and routing, where the solver estimates fidelity–resource trade-offs across transpilation strategies (e.g., different layouts or circuit depths) before actual execution.}


\begin{figure}[t]
	\centering
	\includegraphics[width=0.48\textwidth]{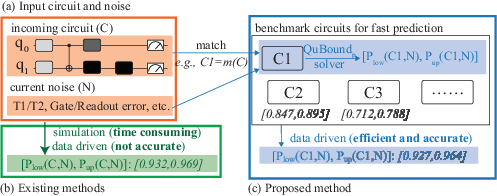}
	\caption{An efficient and accurate $QuBound_p$ solver can provide fidelity estimation for an unseen incoming quantum circuit.}
	\vspace{-8pt}
	\label{fig:usecase}
\end{figure}

\textbf{Comparison with Existing Approaches.}
\jy{Figure \ref{fig:usecase}(a) presents the setting: an incoming circuit $C$ and the current noise profile $N$ (e.g., gate errors). 
The objective is to predict the fidelity of $C$ under $N$.}
\jy{Figure \ref{fig:usecase}(b) illustrates two types of existing approaches. Simulation-based methods directly simulate $C$ under $N$, which is often computationally expensive and time-consuming, and might be inaccurate due to rapidly changing noise in real quantum hardware.}
\jy{Existing data-driven approaches try to learn a global model across various circuits and noise configurations, but they often suffer from high-dimensional input and conflicting labels, reducing accuracy. Even methods such as Linear Regression, which use the same input as our model but without sequential modeling, fail to achieve sufficient accuracy due to limited expressiveness.}

\jy{Figure \ref{fig:usecase}(c) presents our proposed $QuBound_p$ solver.}
Unlike the existing data-driven approach to directly predict fidelity with an extremely large space of possible inputs, the $QuBound_p$ problem significantly reduces the optimization space, targeting one fixed quantum circuit.
It also provides historic noise and performance (obtained by either execution on actual quantum devices or noisy simulation), enabling the use of a chronological neural network for feature extraction.

\textbf{Limitation and Future Works.}
An additional challenge in using the solver for incoming quantum circuits without historical data is that it is not practical to generate all relevant historical information for unseen circuits in real time. However, this does not present a significant barrier, as a potential solution is to build a benchmark pool, as illustrated in the right-hand path of Figure \ref{fig:usecase}.
The key idea is to identify a proxy quantum circuit from a benchmark pool for the incoming circuit $C$.
Then, the proxy circuit, say $C1$ in Figure \ref{fig:usecase}, will be streamed into the $QuBound_p$ solver to obtain the performance bounds.
These bounds will be used to predict the performance/fidelity of circuit $C$.

The above proxy approach for performance prediction is validated through experiments.
The incoming circuit $C$ and $C1$ are matched, with each containing 3 qubits, a circuit length of 13, and 10 quantum gates. However, they differ in gate types and placement.
As shown in Figure \ref{fig:usecase}, the performance of $C1$ is $[0.927,0.964]$, which is very close to $C$ with a performance bound range of $[0.932,0.969]$.
On the other hand, for the unmatched circuit $C2$ with a depth of 38 and 20 gates, its performance bound range is $[0.847,0.895]$.
This indicates that the performance of circuits with similar properties is close.

\jy{While our current proxy matching uses simple heuristics such as qubit count and gate length, we acknowledge that more systematic similarity metrics—such as gate type distributions, connectivity patterns, or circuit depth profiles—may yield more reliable matching. Graph-based representations could support more structured and expressive metrics, capturing both functional and topological similarities.
}

\jy{
For incoming quantum circuits without historical performance traces, there are two possible complementary strategies:}
\jy{(1) As described above, a proxy-based approach can be used, where the incoming circuit is matched to a structurally similar benchmark circuit whose trace is available. The solver can then reuse the proxy's performance trace to estimate the bounds for the incoming circuit.}
\jy{(2) Alternatively, for circuits from the same family (e.g., GHZ), we envision using limited new data to fine-tune the model. This leverages the structural family similarity to improve generalization.
Kindly note that this paper focuses on the development of the $QuBound_p$ solver, the two possible methods will be planned as future work.}



\section{Conclusion}
\label{section8}
In next-generation quantum-centric supercomputing, a quantum performance predictor is essential for fidelity-aware quantum system management (e.g., job allocation and scheduling) and circuit compilation.
This paper proposes a novel data-driven workflow, namely \textit{QuBound}, to accurately predict the bounds of the performance for a given quantum circuit in noisy running environments. 
It designs the performance decomposition component \textit{QuDECOM} to isolate the performance affected by different types of noises and devises a neural network component \textit{QuPRED} to predict the circuit performance for a given noise on one quantum processor.
Experimental results verified the accuracy, efficiency, and scalability of the proposed \textit{QuBound} framework on both noisy simulation and actual IBM quantum computers.
We envision that \textit{QuBound} could effectively provide a useful metric for optimizations in the incoming quantum-centric supercomputing. 

\bibliographystyle{IEEEtran}
\bibliography{bib(reduce)}
\vspace{-15mm}
\begin{IEEEbiography}[{\includegraphics[width=1in,height=1.25in,clip,keepaspectratio]{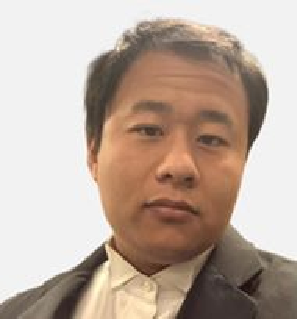}}]{Jinyang Li} is currently a Ph.D. student in the Department of Electrical and Computer Engineering at George Mason University. He holds a bachelor’s and a master’s degree in computer science from Virginia Tech. His research focuses on quantum computing, with a particular emphasis on unstable quantum noise, variational quantum algorithms, and quantum learning tasks. He has published several papers in conferences in the field.
\end{IEEEbiography}
\vspace{-15mm}
\begin{IEEEbiography}[{\includegraphics[width=1in,height=1.25in,clip,keepaspectratio]{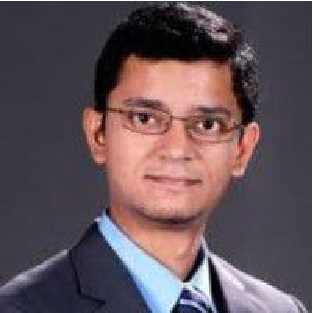}}]{Samudra Dasgupta}
is a researcher at the Quantum Science Center, Oak Ridge National Laboratory. He earned his Ph.D. in Data Science, specializing in Quantum Computing, from the Bredesen Center at the University of Tennessee, Knoxville, in partnership with ORNL. His research explores the reliability challenges of noisy quantum computations. Samudra also holds an M.S. in Applied Physics from Harvard University and a B.Tech. in Electronics and Electrical Communications Engineering from the Indian Institute of Technology, Kharagpur.
\end{IEEEbiography}
\vspace{-15mm}
\begin{IEEEbiography}[{\includegraphics[width=1in,height=1.25in,clip,keepaspectratio]{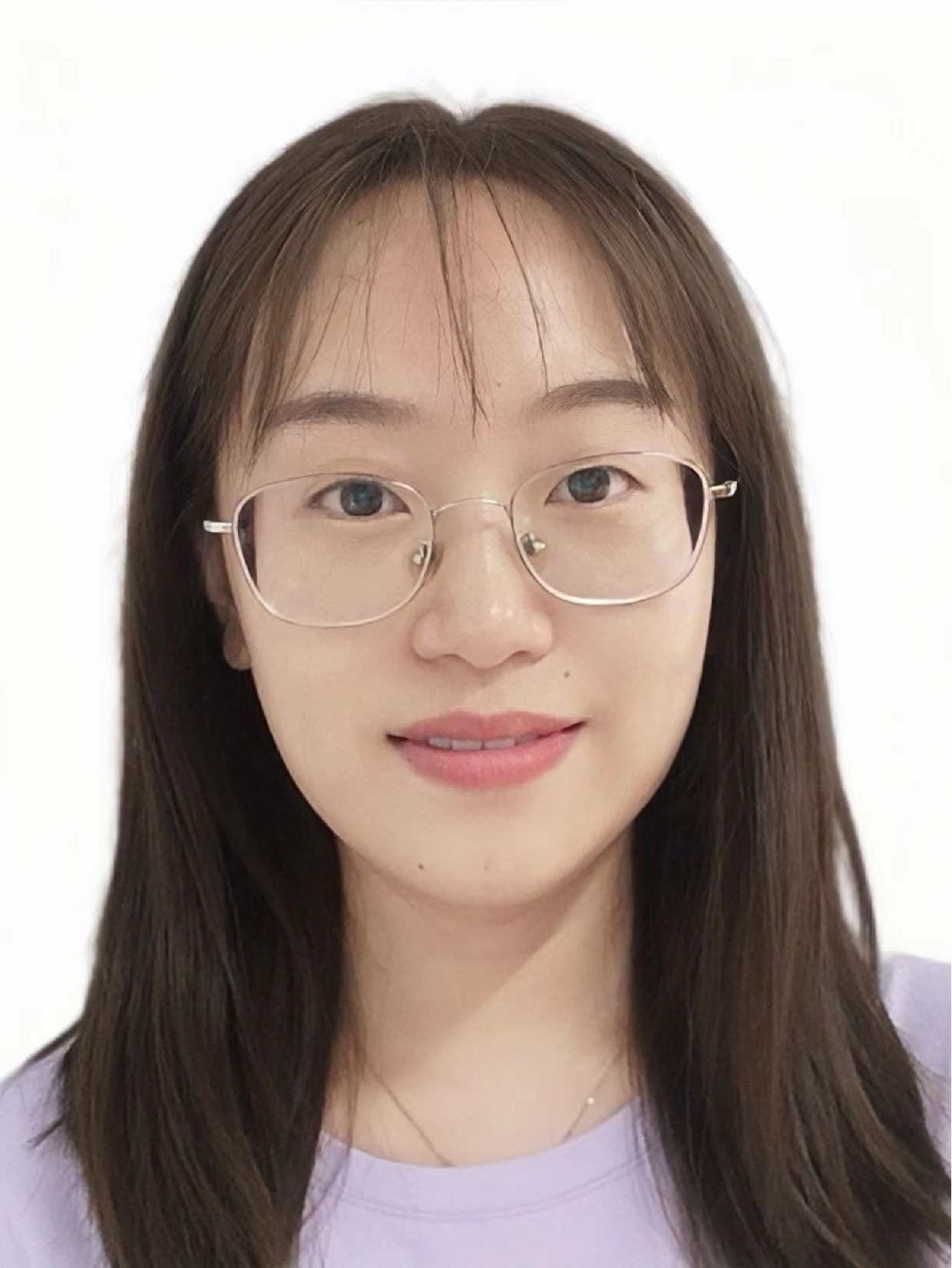}}]{Yuhong Song} is currently a postdoctoral research fellow in the Department of Electrical and Computer Engineering at George Mason University. She received her Ph.D. degree from the School of Computer Science and Technology, East China Normal University, in 2024. Previously, she was a visitor scholar in the Department of Computer Science and Engineering at The Chinese University of Hong Kong. Her current research interests include quantum computing, embedded systems, and software-hardware co-design.
\end{IEEEbiography}
\vspace{-15mm}
\begin{IEEEbiography}[{\includegraphics[width=1in,height=1.25in,clip,keepaspectratio]{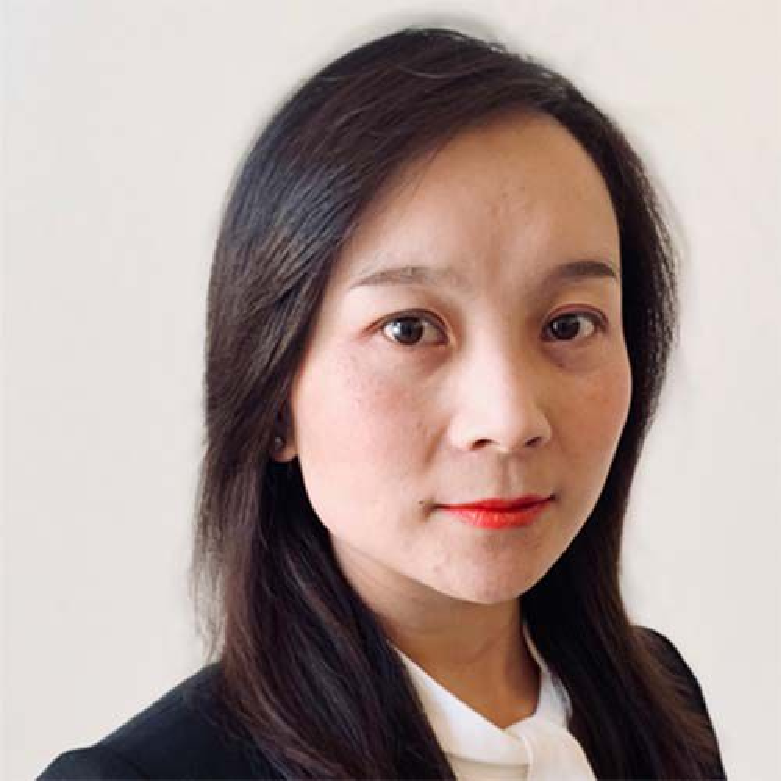}}]{Lei Yang} is currently an Assistant Professor in the Department of Information Sciences and Technology at George Mason University. 
Lei’s primary research interests lie in the joint area of Hardware/Software Co-Exploration for Neural Network Architectures, Embedded Systems, and High-Performance Computing. 
She has published more than 50 research articles in premier international conferences and journals, including DAC, CODES+ISSS, ASP-DAC, ICCD, HPCC, RTCSA, IEEE Transactions (TC, TPDS, TVLSI, TCAD), ACM Transactions (TECS).
\end{IEEEbiography}
\vspace{-15mm}
\begin{IEEEbiography}[{\includegraphics[width=1in,height=1.25in,clip,keepaspectratio]{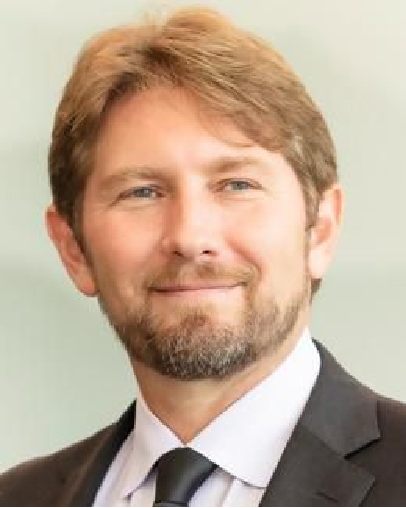}}]{Travis Humble}
is the director of the Quantum Science Center and a Distinguished Scientist at Oak Ridge National Laboratory, where he leads efforts in advancing quantum technologies and infrastructure to support the Department of Energy’s mission of scientific discovery. As director of both the Quantum Science Center and ORNL’s Quantum Computing User Program, he oversees the development, management, and benchmarking of quantum computing platforms for diverse scientific applications. Travis is also editor-in-chief of ACM Transactions on Quantum Computing and holds a joint faculty appointment with the University of Tennessee’s Bredesen Center. 
\end{IEEEbiography}
\vspace{-15mm}
\begin{IEEEbiography}[{\includegraphics[width=1in,height=1.25in,clip,keepaspectratio]{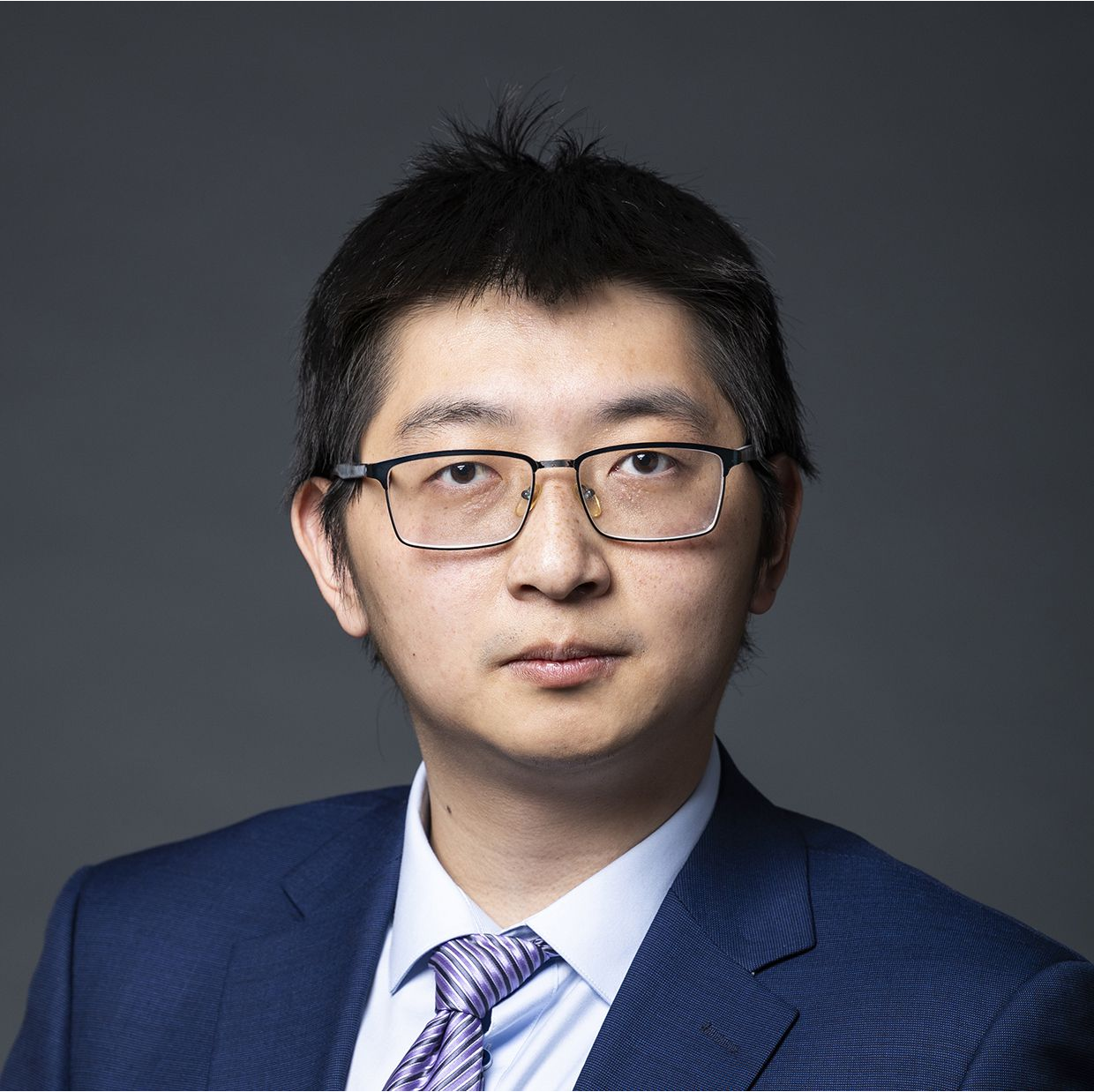}}]{Weiwen Jiang}is an Assistant Professor in the ECE department at George Mason University. His research focuses on quantum computing design automation. He served as the Technical Track Co-Chair at IEEE QuantumWeek (2023-2025). He created the Quantum System Stability and Reproducibility Workshop (StableQ). He is the recipient of the NSF CAREER Award 2025, GMU's Presidential Award for Faculty Excellence in Research 2025, and ACM Sigda Meritorious Service Award 2024. His works have won Best Paper Awards in IEEE QuanutumWeek'23, TCAD'21, ICCD'17, and NVMSA'15. He was the technique program track chair of IEEE QuantumWeek 2023 and 2024.
\end{IEEEbiography}

\end{document}